\documentclass[showpacs,prb,floatfix,twocolumn,amsmath,a4]{revtex4}
\usepackage{graphicx}
\usepackage{amsmath}
\usepackage{amssymb}
\usepackage{bm}
\usepackage{dcolumn}
\usepackage{subfigure}

\begin{document}
\title{First-principles study of the interaction and charge transfer between graphene and metals}
\author{P. A. Khomyakov,$^1$ G. Giovannetti,$^{1,2}$ P. C. Rusu,$^1$ G. Brocks,$^1$
J. van den Brink,$^{2,3}$ and P. J. Kelly$^1$} \affiliation{$^1$
Faculty of Science and Technology and MESA$^+$ Institute for
Nanotechnology, University of Twente, P.O. Box 217, 7500 AE Enschede, The Netherlands\\
$^2$Instituut-Lorentz for Theoretical Physics,
Universiteit Leiden, P.O. Box 9506, 2300 RA Leiden, The Netherlands\\
$^{3}$ Institute for Molecules and Materials, Radboud Universiteit,
Heyendaalseweg 135, 6525 AJ Nijmegen, The Netherlands }

\begin{abstract}
Measuring the transport of electrons through a graphene sheet
necessarily involves contacting it with metal electrodes. We study the
adsorption of graphene on metal substrates using first-principles
calculations at the level of density functional theory. The bonding of
graphene to Al, Ag, Cu, Au and Pt(111) surfaces is so weak that its
unique ``ultrarelativistic'' electronic structure is preserved. The
interaction does, however, lead to a charge transfer that shifts the
Fermi level by up to 0.5 eV with respect to the conical points. The
crossover from $p$-type to $n$-type doping occurs for a metal with a
work function $\sim 5.4$ eV, a value much larger than the work function
of free-standing graphene, 4.5 eV. We develop a simple analytical model
that describes the Fermi level shift in graphene in terms of the metal
substrate work function. Graphene interacts with and binds more
strongly to Co, Ni, Pd and Ti. This chemisorption involves
hybridization between graphene $p_z$-states and metal $d$-states that
opens a band gap in graphene. The graphene work function is as a result
reduced considerably. In a current-in-plane device geometry this should
lead to $n$-type doping of graphene.
\end{abstract}
\date{\today}
\pacs{73.63.-b, 73.20.Hb, 73.40.Ns, 81.05.Uw}
\maketitle

\section{Introduction}
The history of carbon-based electronics begins with the discovery of
fullerenes and carbon nanotubes that are zero- and one-dimensional, respectively.\cite{Kroto:nat85,Iijima:nat91}
A very important recent development was the preparation of single
monolayers of graphite, now more commonly called graphene, on
insulating substrates using micromechanical cleavage
\cite{Novoselov:pnas05} that has made possible electron transport
experiments on this purely two-dimensional system.
\cite{Novoselov:sc04,Novoselov:nat05,Zhang:nat05,Zhou:natp06,Bolotin:ssc08,Danneau:prl08}
These transport measurements reveal high charge carrier mobilities,
quantization of the conductivity, and a zero-energy anomaly in the
quantum Hall effect, as predicted theoretically.
\cite{Shon:jpsj98,Ando:jpsj02,Gusynin:prl05,Katsnelson:natp06,vandenBrink:natn07}
The theoretical studies explain these spectacular effects in terms of
graphene's unique electronic structure. Although a single graphene
sheet is a zero-gap semiconductor with a vanishing density of states at
the Fermi energy, it shows metallic behavior due to topological
singularities at the $K$-points in the Brillouin
zone,\cite{Shon:jpsj98,Ando:jpsj02} where the conduction and valence
bands touch in so-called conical or Dirac points and the dispersion is
essentially linear within $\pm 1$ eV of the Fermi energy. Its high
charge carrier mobility and peculiar electronic properties have
stimulated considerable research into the possibilities of using
graphene for electronic and spintronic applications.

Graphene is often treated theoretically as a free-standing
two-dimensional sheet. Though this often appears to be a reasonable
model for describing observed properties, in many experimental
situations there is some deviation from this ideal because there is
some form of physical contact with the environment. This can consist of
atomic and molecular impurities in or on the graphene sheet, contact
with an insulating substrate, a gate electrode or metallic leads, etc.
\cite{Novoselov:nat05,Bolotin:ssc08,Sabio:prb08,Boukhvalov:prb08,Lee:natn08}
While the Fermi energy of free-standing graphene coincides with the
conical points, adsorption on substrates can alter its electronic
properties
significantly.\cite{Oshima:jpcm97,Dedkov:prb01,Bertoni:prb05,
NDiaye:prl06,Karpan:prl07,Giovannetti:prb07,Marchini:prb07,Uchoa:prb08,
Rotenberg:natm08,Wu:prl08,Giovannetti:prl08} For example, the weak
interaction of graphene adsorbed on the (0001) surface of insulating
hexagonal boron nitride (h-BN) is enough to destroy graphene's
characteristic conical points and open a band gap of some 50
meV.\cite{Giovannetti:prb07} Even when the interaction is sufficiently
weak to leave the conical points essentially unchanged, it can lead to
a large shift of the Fermi energy away from the conical
points.\cite{Giovannetti:prb07,Giovannetti:prl08}

\begin{figure*}[!tbp]
\begin{center}
\includegraphics[width=1.5\columnwidth,angle=0]{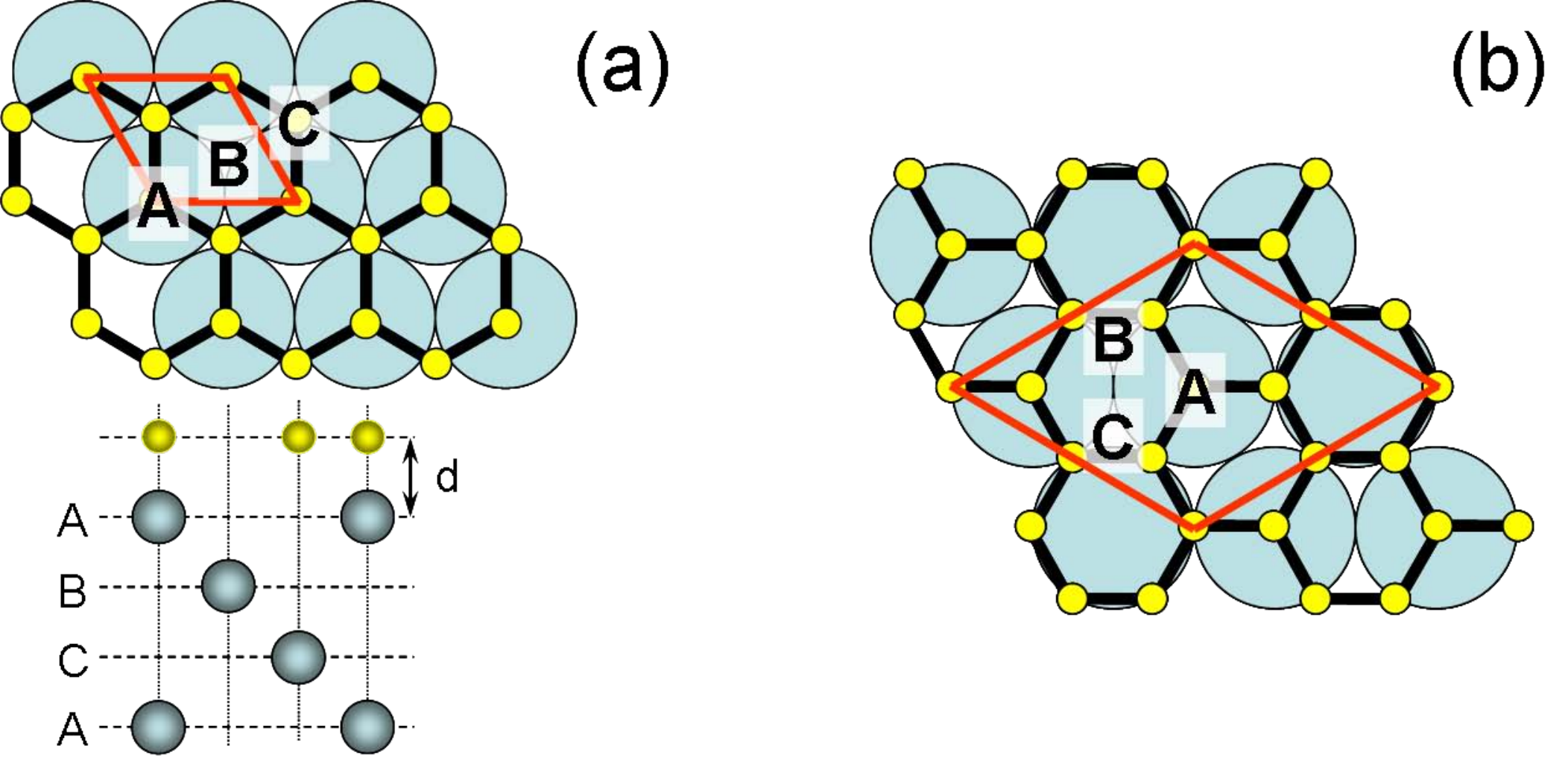}
\caption{(Color online) (a) The most stable symmetric configuration of
graphene on Cu,  Ni and Co (111) has one carbon atom on top of a metal
atom (A site), and the second carbon on a hollow site (C site). (b)
Graphene on Al, Au, Pd and Pt(111) can be modeled in a $2\times 2$
graphene supercell with 8 carbon atoms and 3 metal atoms per layer.
Shown is the most stable symmetric geometry.}
\label{ref:fig1}
\end{center}
\end{figure*}

Since measurement of the electronic transport properties of graphene
requires making contacts with metal leads,
\cite{Novoselov:nat05,Lee:natn08,Karpan:prl07,Schomerus:prb07,Blanter:prb07,Huard:prb08,Nouchi:apl08,
Russo:condmat09} it is
important to understand such electronic and structural properties as
the charge transfer between graphene and the metal substrate, the
graphene-metal binding energies, distances etc. Charge transfer at a
metal-graphene interface results in doping of the graphene sheet.
Because the sign and the magnitude of the doping depend upon the metal,
$p$-$n$ junctions can be realized by attaching electrodes of different
metals to graphene.
\cite{Cheianov:prb06,Huard:prl07,Ozyilmaz:prl07,Williams:sc07,Tworzydlo:prb07,
Fogler:prb08,Park:natp08,Lee:natn08,Gorbachev:nanol08}

There have been numerous theoretical and experimental studies on
semiconducting carbon nanotubes contacted to metals such as Al, Au, Pt,
Pd, Ca and Ti.\cite{Okada:prl05,Nosho:apl05,Zhu:apl06,Meng:jap07,Vitale:jacs08}
Since a graphene sheet can be considered as a carbon nanotube of
infinite radius, the chemical interaction between graphene and metal
substrates can be expected to be similar to that between metal contacts
and nanotubes.

In this paper we use first-principles calculations at the level of
density functional theory (DFT) to characterize the adsorption of
graphene on a variety of metal substrates. A preliminary account of our
results was given in Ref.~\onlinecite{Giovannetti:prl08}. The (111)
surfaces of Al, Co, Ni, Cu, Pd, Ag, Pt, Au and the Ti(0001) surface
cover a wide range of work functions and different types of chemical
bonding, which allows for a systematic study of the metal-graphene
interface. We focus on the interaction and charge transfer between
graphene and the metal substrate, and in particular the effects they
have on the doping of graphene by the metal. Because the charge
redistribution at the graphene-metal interface can be characterized
experimentally by measuring the work function of the graphene-covered
metal, we also calculate the work functions of these systems.

The structural details of the metal-graphene interfaces will be
presented elsewhere.\cite{Khomyakov:unpub1} The most important result
for the purposes of this study is that there are two classes of
graphene-metal interfaces. Whereas graphene is chemisorbed on Co, Ni,
Pd and Ti, the binding to Al, Cu, Ag, Au and Pt is much weaker. The
electronic structure of graphene is strongly perturbed by chemisorption
but is essentially preserved in the weak binding ``physisorption''
regime. For physisorbed graphene there is generally electron transfer
to (from) the metal substrate, causing the Fermi level to move downward
(upward) from the graphene conical points. This can be viewed as doping
graphene with holes (electrons) by adsorption.

Naively one might expect the type and amount of doping to depend only
on the difference between the work functions of free-standing graphene
and of the clean metal surface. At typical equilibrium separations, the
potential profile and therefore the doping are, however, altered
significantly by an interface dipole arising from a direct short-range
metal-graphene interaction. Using the DFT results, we develop an
analytical model that quantitatively describes the doping of
physisorbed graphene. This model also predicts how physisorption of
graphene modifies the metal work function.

In order to characterize the doping of chemisorbed graphene a different
approach must be used. Since chemisorption perturbs the electronic
structure of graphene strongly, doping cannot be simply deduced from
the shift of the Fermi level with respect to the conical points.
Instead, we consider the work function of the graphene-covered metal,
which is always a well-defined quantity. In a current-in-plane
transport experiment only part of the graphene sheet covers (or is
covered by) the metal electrode, whereas an adjacent part is
free-standing. The difference between the work function of the
graphene-covered metal electrode and free-standing graphene then
determines the direction of the charge transfer between these two parts
and hence the doping. According to this model, graphene is doped
$n$-type by Co, Ni, Pd and Ti contacts.

\begin{table*}[!tpb]
\caption{$a_{\rm hex}^{\rm exp}$ and $\tilde{a}_{\rm hex}^{\rm exp}$
represent the experimental cell parameters of the surface unit cells
shown in Fig.~\ref{ref:fig1}(a) and Fig.~\ref{ref:fig1}(b),
respectively, for graphene on various metals. All calculations are
performed with the lattice constant of graphene optimized using the
LDA, $a_{\rm hex}=2.445$ \AA. The calculated equilibrium separation
$d_{\rm eq}$ is the separation in the $z$ direction between the carbon
atoms of the graphene sheet and the relaxed positions of the topmost
metal layer, averaged where applicable over the carbon and metal atoms
in the lateral supercell. The binding energy $\Delta E$ is the energy
per carbon atom required to remove the graphene sheet from the metal
surface. $W_{\rm M}$ and $W$ are, respectively, the calculated work
functions of the clean metal surfaces and of free-standing and adsorbed
graphene, and $W_{\rm M}^{\rm exp}$, $W^{\rm exp}$ are the
corresponding experimental values. $\Delta E_{\rm F}$ is the Fermi
level shift of physisorbed graphene.}
\begin{ruledtabular}
\begin{tabular}{lllllllllll}
                      & Gr   & Ti   &  Ni   & Co   & Pd   & Al   &  Ag  & Cu   & Au   & Pt     \\
\hline
 $a_{\rm hex}^{\rm exp}$ (\AA)
                      & 2.46 & 2.95 & 2.49  & 2.51 &      &      &      & 2.56 &      &      \\
 $\tilde{a}_{\rm hex}^{\rm exp}$ (\AA)
                      & 4.92 &      &       &      & 4.76 & 4.96 & 5.00 &      & 4.99 & 4.81 \\
 ${d_{\rm eq}}$ (\AA) &      & 2.1  & 2.05  & 2.05 & 2.30 & 3.41 & 3.33 & 3.26 & 3.31 & 3.30  \\
 ${\Delta}E$ (eV)    &      & 0.180  & 0.125   &  0.160 & 0.084   &  0.027  &  0.043  & 0.033   & 0.030   & 0.038    \\
 $W_{\rm M}$ (eV)     &      & 4.70 & 5.47  & 5.44 & 5.67 & 4.22 & 4.92 & 5.22 & 5.54 & 6.13  \\
 $W_{\rm M}^{\rm exp}$
 (eV)                 &     & 4.58\footnote{Ref.~\onlinecite{Jonker:prb81}} & 5.35\footnote{Ref.~\onlinecite{Michaelson:jap77}} & 5.55\footnote{Ref.~\onlinecite{Vaara:ss98}} & 5.6$^{b}$ & 4.24$^{b}$ & 4.74$^{b}$ & 4.98$^{b}$ & 5.31$^{b}$ &  6.1\footnote{Ref.~\onlinecite{Derry:prb89}} \\
$W$ (eV)             & 4.48 & 4.14 & 3.66  & 3.78 & 4.03 & 4.04 & 4.24 & 4.40 & 4.74 & 4.87  \\
$W^{\rm exp}$ (eV)   & 4.6\footnote{Ref.~\onlinecite{Oshima:jpcm97}} &      & 3.9$^{e}$   &      & 4.3$^{e}$  &      &      &      &      & 4.8$^{e}$ \\
$\Delta E_{\rm F}$ (eV)  &      &       &       &       &      & $-0.57$ & $-0.32$ & $-0.17$ & 0.19 & 0.33
\label{ref:tab1}
\end{tabular}
\end{ruledtabular}
\end{table*}

The paper is organized as follows. In Sec.~\ref{sec:compdetails} we
state the most important computational details of the density
functional calculations and summarize in Sec.~\ref{sec:mgbinding} the
key results of a more extensive study of the binding of graphene to
various metal substrates.\cite{Khomyakov:unpub1}
Section~\ref{subsec:abinitio} contains results of the first-principles
calculations for the doping and work function of graphene adsorbed on
these different substrates. A phenomenological model to describe the
doping and work function of physisorbed graphene is introduced in
Sec.~\ref{subsec:wmgbinding} and in Sec.~\ref{subsec:smgbinding}
chemisorbed graphene is discussed. The sensitivity of the results to
the computational approximations used is discussed in
Sec.~\ref{sec:functional}. A short discussion and conclusions are
presented in Sec. \ref{sec:conclusions}.

\section{Computational details}
\label{sec:compdetails} We calculate DFT ground state energies and
optimized geometries using a plane wave basis set and the PAW formalism
at the level of the local (spin) density approximation,
L(S)DA,\cite{Perdew:prb81} as implemented in the VASP
code.\cite{Blochl:prb94b,Kresse:prb99,Kresse:prb93,Kresse:prb96} The
plane wave kinetic energy cutoff is set at 400 eV. A metal surface is
modelled in a supercell as a finite number of layers of metal plus a
region of vacuum repeated periodically in the direction perpendicular
to the layers. The supercell used to model the graphene metal
adsorption is constructed from a slab of six layers of metal atoms with
a graphene sheet adsorbed on one side and a vacuum region of $\sim 12$
\AA. A dipole correction is applied to avoid spurious interactions
between periodic images of the slab.\cite{Neugebauer:prb92}

We choose the in-plane lattice constant of graphene equal to its
optimized LDA value, $a = 2.445$ \AA, adapting the lattice constants of
the metals accordingly. The graphene honeycomb lattice then matches the
triangular lattice of the metal (111) surfaces in the lateral unit
cells shown in Fig.~\ref{ref:fig1}. The approximation made by this
matching procedure is reasonable since the mismatch with the lattice
parameters of the metal (111) surfaces is only 0.8-3.8\%, as seen in
Table~\ref{ref:tab1}. In optimizing the geometry, the positions of the
carbon atoms as well as those of the top two layers of metal atoms are
allowed to relax. All results reported in this paper are obtained for
structures adapted to the LDA optimized in-plane lattice constant of
graphene.

We use the tetrahedron scheme \cite{Blochl:prb94a} for accurate
Brillouin Zone (BZ) integrations, sampling the BZ of the small and
large cells in Fig.~\ref{ref:fig1} with $36\times 36$ and $24\times 24$
\textbf{k}-point grids, respectively, and explicitly including the
$\Gamma $, $K$ and $M$ high symmetry points. Note that on doubling the
graphene lattice vectors to match those of Au, Pt, Cu, Ag, Al and Pd,
the $K$ point corresponding to the primitive unit cell of graphene is
folded down onto the $\bar{K}$ point of the smaller Brillouin zone. The
electronic self-consistency criterion is set to $10^{-7}$eV. Such a
strict convergence is required to obtain accurate forces, which are
essential in order to obtain reliable optimized structures. Total
energies are converged to within $10^{-6}$~eV in respect of ionic
relaxation. Explicit total energy calculations show that the structures
in Fig.~\ref{ref:fig1} represent the most stable symmetric
configurations of graphene on the metal substrates studied, in
agreement with experimental results where available.
\cite{Oshima:jpcm97}

Detailed interfaces structures will be reported elsewhere.
\cite{Khomyakov:unpub1} Here we note that the L(S)DA functional gives a
much better description of graphene-metal substrate binding energies
and equilibrium distances than the commonly used generalized gradient
approximation (GGA) functionals. Since the work functions, calculated
with the L(S)DA, of clean metal surfaces and of those covered with
graphene are sufficiently accurate, we use the L(S)DA functional. In
Sec.~\ref{sec:conclusions} we will show that, provided the
graphene-metal substrate equilibrium separation is obtained correctly,
the charge transfer and consequently the doping of graphene do not
depend strongly on the choice of density functional.
\begin{figure*}[!tpb]
\begin{center}
\includegraphics[width=2.0\columnwidth]{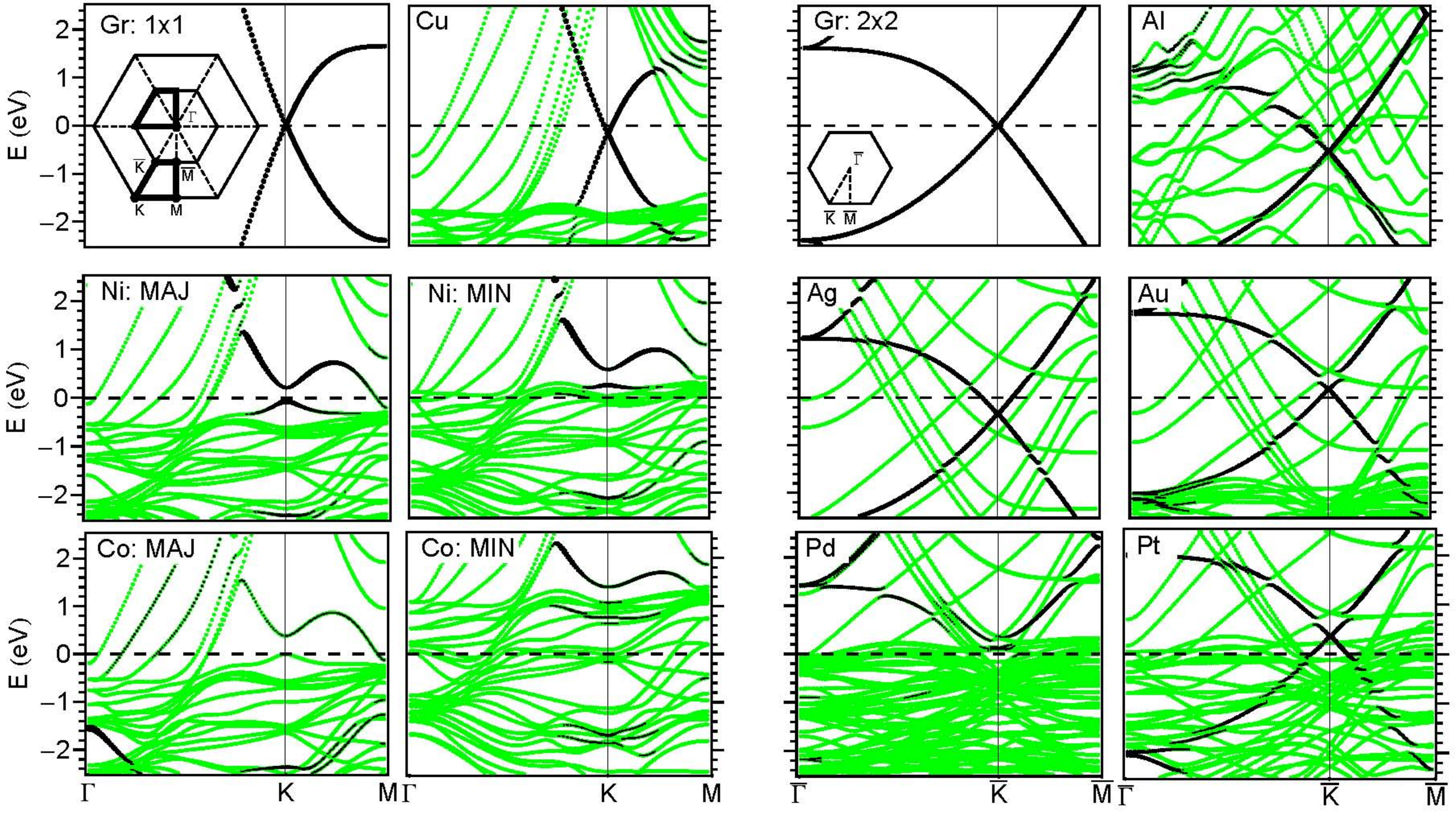}
\end{center}
\caption{(Color online) Band structures of graphene adsorbed upon Au,
Pt, Cu, Ag, Al, Pd, Ni and Co (111) substrates. The Fermi level is at
zero energy. The amount of carbon $p_{z}$ character is indicated by the
blackness of the bands. The conical point corresponds to the crossing
of bands at $K$ with predominantly $p_{z}$ character, as is clearly
visible for (physisorbed) graphene on Au, Pt, Cu, Ag, and Al. For
(chemisorbed) graphene on Pd, Ni, and Co, the conical points disappear
and the bands have a mixed character. The labels MIN/MAJ indicate the
majority and minority spin bands of graphene on Ni and Co. The first and third top panels correspond to the band
structure of free-standing graphene calculated with the primitive unit cell and $2\times 2$ graphene supercell, respectively.
Inset: the two-dimensional Brillouin zones of graphene for the primitive unit cell and $2\times 2$ graphene supercell. In the supercell the bands are downfolded, the area enclosed by the bold lines in the primitive BZ translates to the corresponding one in the supercell BZ.
} \label{ref:fig2}
\end{figure*}

Matching the graphene lattice with the Ti(0001) surface is more
difficult since there is a lattice mismatch of 20\%. To accommodate
this mismatch we use a graphene $7\times 7$ lateral supercell and a BZ
sampling of a similar density as above. The equilibrium separation
given in Table~\ref{ref:tab1} is the value obtained by averaging over
the lateral supercell. The details of the graphene/Ti(0001)
calculations will be reported elsewhere. Here we focus on the charge
redistribution at the interface and the doping of graphene.

\section{Results}
\subsection{Metal-graphene binding}
\label{sec:mgbinding} The calculated equilibrium bonding distances, the
binding energies and the work functions for adsorption of graphene on
all metal substrates studied in this paper are listed in
Table~\ref{ref:tab1}. The binding energies $\Delta E$ and equilibrium
separations $d_{\rm eq}$ immediately show that the metals can be
divided into two classes. For graphene adsorbed on Co, Ni, Pd(111) and
Ti(0001), $\Delta E\gtrsim 0.1$ eV/carbon atom and $d_{\rm eq}\lesssim
2.3$ \AA. In contrast, adsorption on Al, Cu, Ag, Au and Pt(111) leads
to much weaker bonding, $\Delta E \lesssim 0.04$ eV/carbon atom, and
larger equilibrium separations, $d_{\rm eq} \sim 3.3$ \AA. The
equilibrium geometries and distances obtained are in agreement with
available experimental data and calculations
\cite{Oshima:jpcm97,Gamo:ss97,Bertoni:prb05,Qi:ss05} and appear to be
similar to the bonding found between graphene and carbon nanotubes;
carbon nanotubes are usually bonded strongly to Pd and Ti whereas the
bonding with Al, Ag, Au, Ca and Pt is weaker.
\cite{Okada:prl05,Nosho:apl05,Zhu:apl06,Meng:jap07,Vitale:jacs08}

The difference between the two classes of metal substrates is reflected
in the electronic structure of adsorbed graphene as shown in
Fig.~\ref{ref:fig2}. When the binding energy is large, i.e., if
graphene is adsorbed on Co, Ni, Pd or Ti, the graphene bands are
strongly perturbed. In particular, the characteristic conical points of
graphene at $K$ are destroyed. Graphene $p_z$-states hybridize strongly
with the metal $d$-states and the corresponding bands acquire a mixed
graphene-metal character. It demonstrates that graphene is chemisorbed
on these substrates.

In contrast, if the metal-graphene interaction is weaker, i.e., when
graphene is adsorbed on Al, Cu, Ag, Au or Pt, the graphene bands,
including their conical points at $K$, can still be clearly identified.
We reserve the term physisorption to describe this type of bonding.
Unlike in the case of free-standing graphene where the Fermi level
coincides with the conical point, physisorption generally shifts the
Fermi level. Even when there is no interaction or the interaction is
weak, this does not preclude the transfer or charge between graphene
and the metal substrate resulting from the equilibration of the
chemical potentials.

\subsection{Doping of physisorbed graphene}
\label{subsec:abinitio}

\begin{figure}[!tpb]
\includegraphics[width=1.0\columnwidth]{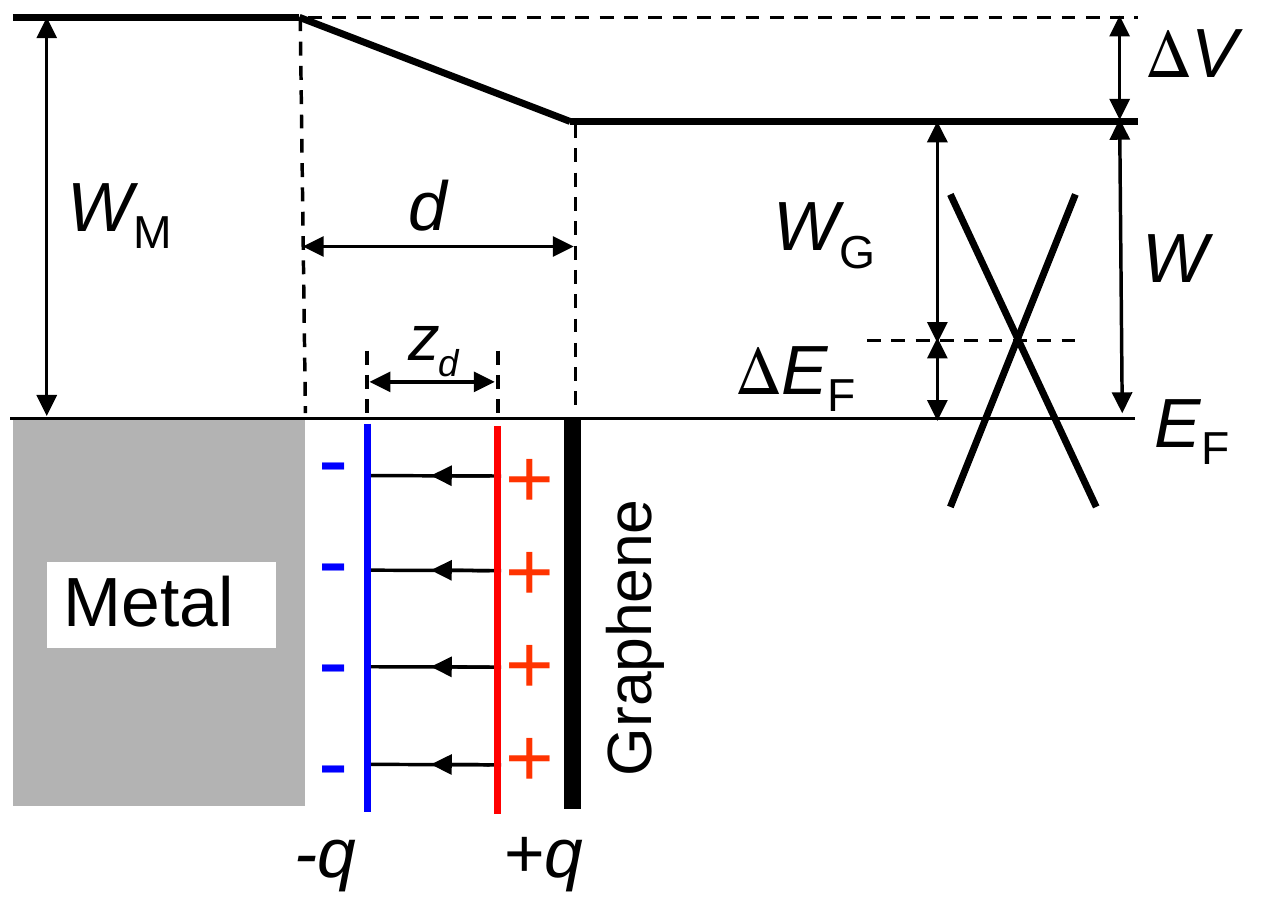}
\caption{(Color online) Schematic illustration of the parameters used
in modelling the interface dipole and potential step formation at the
graphene-metal interface.} \label{ref:fig4a}
\end{figure}
\begin{figure}[tbp]
\begin{center}
\includegraphics[width=1.0\columnwidth]{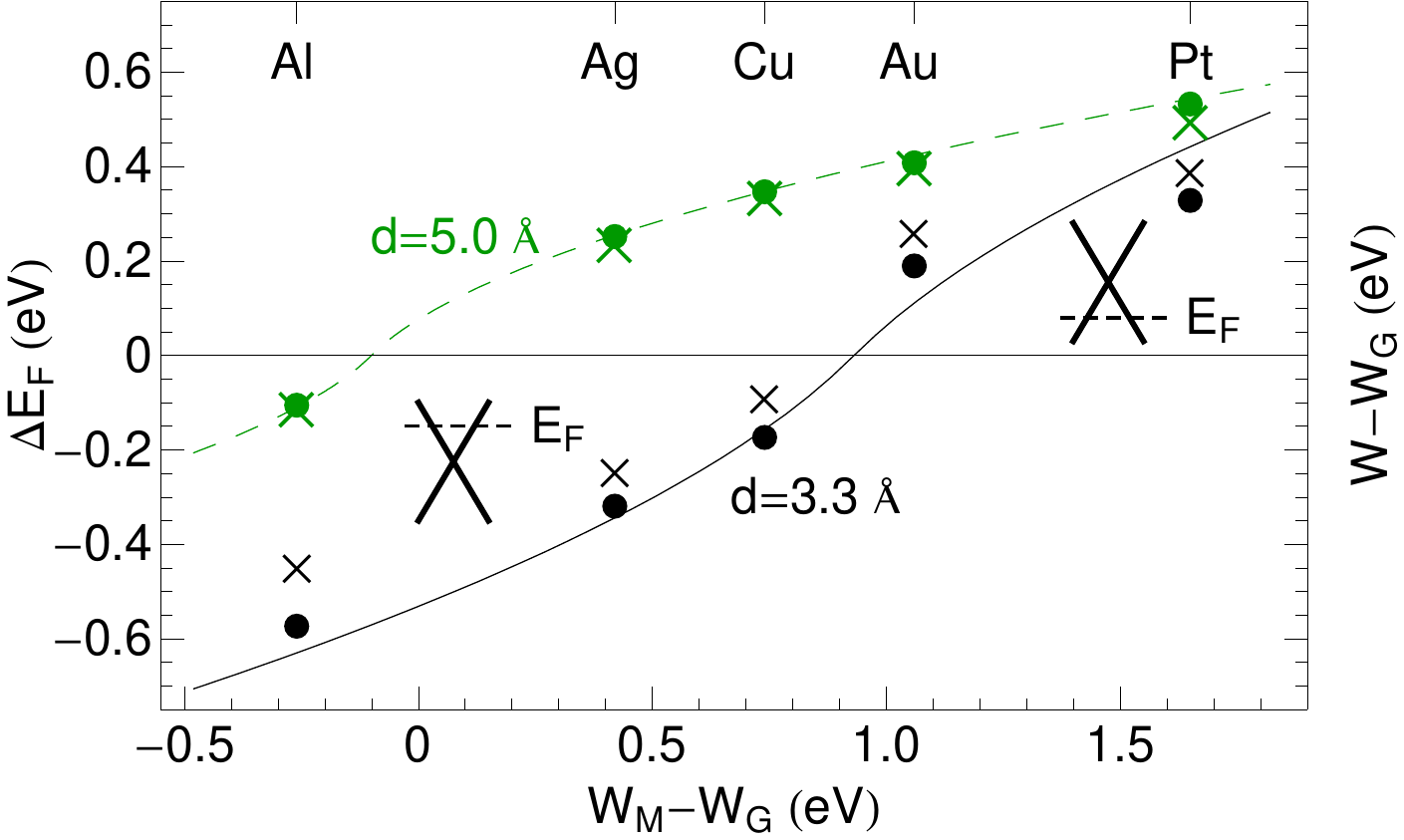}
\caption{(Color online) Calculated Fermi energy shift with respect to
the conical point, $\Delta E_{\rm F}$ (dots), and $W - W_{\rm G}$
(crosses) as a function of the clean metal-graphene work function
difference $W_{\rm M} - W_{\rm G}$.
The lower (black) and
the upper (green/grey) points are for the equilibrium ($\sim 3.3$ \AA)
and large (5.0 \AA) graphene-metal surface distances, respectively. The
solid and dashed lines follow from the model of Eq.~(\ref{eqn1}). The
insets illustrate the position of the Fermi level with respect to the
conical point.} \label{ref:fig3}
\end{center}
\end{figure}

In physisorbed graphene the conical points in the graphene band
structure are preserved, but charge transfer to or from the metal
substrate shifts the Fermi level. A schematic representation of the
parameters we use to describe this situation is shown in
Fig.~\ref{ref:fig4a} for the case of electron transfer from graphene to
the metal. A shift upwards (downwards) with respect to the conical
points means that electrons (holes) are donated by the metal to
graphene, making the latter $n$-type ($p$-type) doped. We extract the
Fermi level shifts $\Delta E_\mathrm{F}$ of graphene physisorbed on a
number of metals from the band structures shown in Fig.~\ref{ref:fig2}
and plot them in Fig.~\ref{ref:fig3}. At equilibrium separations from
the metal substrates, graphene is doped $n$-type on Al, Ag and Cu, and
$p$-type on Au and Pt. In the following section we develop a
phenomenological model to describe these first-principles results. In
the remaining part of this section we identify the physical parameters
that play a role in this model.

The work function $W$ of a graphene-covered metal is given by the
position of the Fermi level ($W = - E_\mathrm{F}$). Because the density
of states of graphene is so small compared to that of the local density
of states of a typical transition metal surface, the shifts required to
equilibrate the Fermi levels when charge transfer occurs take place
almost entirely in graphene. For physisorbed graphene where the
interaction is so weak that its electronic structure is unchanged, $W$
should be related to the Fermi level shift in a simple way
\begin{equation}
\label{eq1}
\Delta E_\mathrm{F} = W - W_\mathrm{G},
\end{equation}
where $W_\mathrm{G}$ is the work function of free-standing graphene.
The work function shifts are calculated separately and are plotted in
Fig.~\ref{ref:fig3}. We see that while Eq.~(\ref{eq1}) holds for a
relatively large separation $d=5.0$ \AA\ between the graphene sheet and
the metal surface, there is a small deviation of $\sim 0.08$~eV at the
equilibrium separation $d\approx3.3$~\AA\ which can be traced to a
perturbation of the graphene electronic structure by physisorption,
that cannot be described as a rigid shift. In the following discussions
we will neglect this small (non-rigid shift) perturbation. When
graphene is chemisorbed and the Fermi level shift cannot be determined
from the strongly perturbed band structure, the work function $W$ is
still a well-defined parameter.

\begin{figure}[!tpb]
\includegraphics[width=1.0\columnwidth]{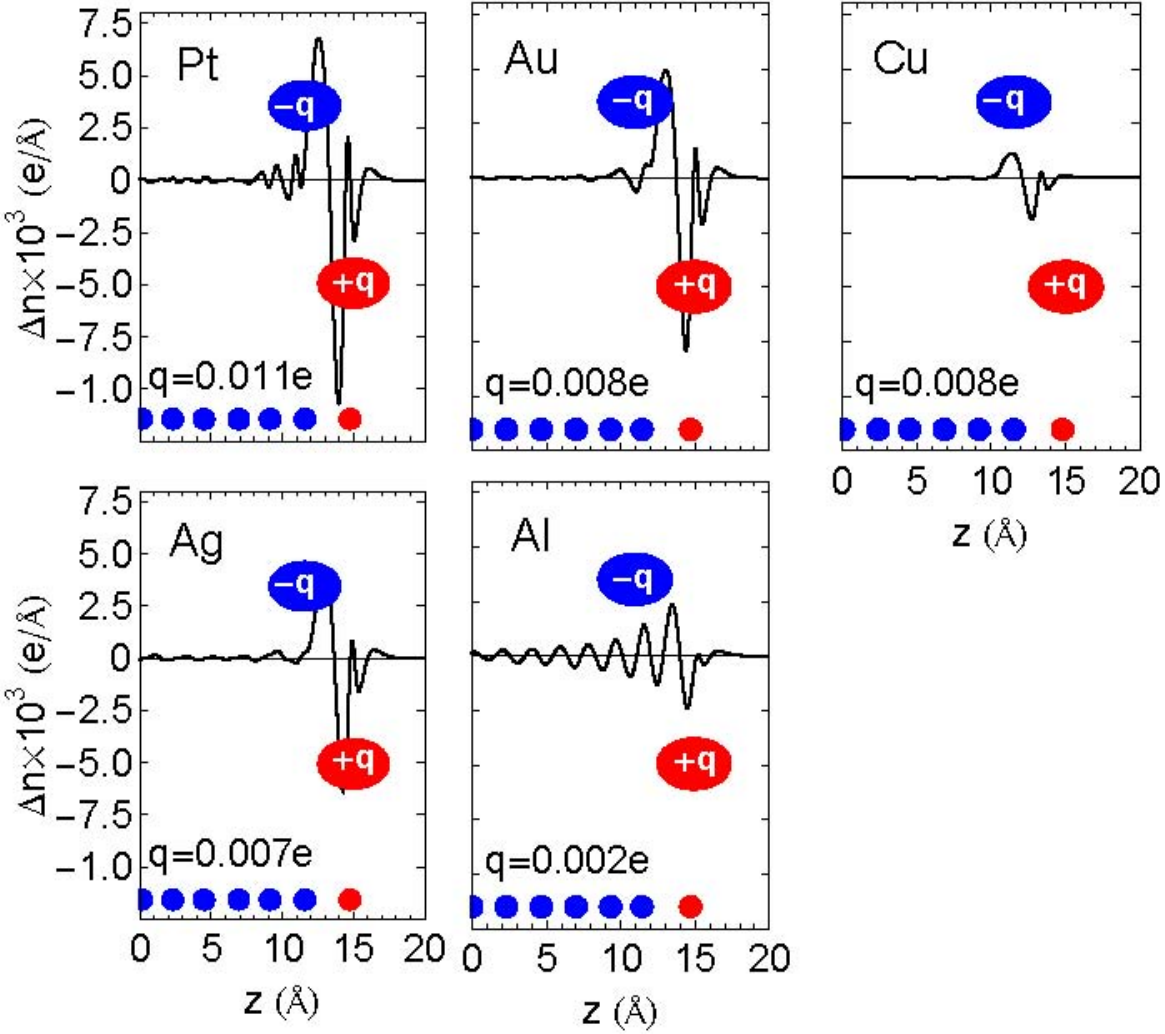}
\caption{(Color online) Plane-averaged electron difference density
$\Delta n (z)$ (per unit cell) showing the charge displacement upon
physisorption of graphene on M(111) surfaces where M = Al, Ag, Cu, Au,
and Pt. $q/e$ is the number of electrons per carbon atom calculated by
integrating $\Delta n (z)$ from the central node to infinity.
\cite{footnote1}
} \label{ref:fig4b}
\end{figure}

Because the work functions of graphene, $W_{\rm G}$, and of most metal
surfaces, $W_{\rm M}$, differ, electrons are transferred from one to
the other to equilibrate the Fermi levels if the two systems
communicate. Charge transfer between metal and graphene results in the
formation of an interface dipole layer and its associated potential
step, $\Delta V$.  We can use the plane-averaged electron densities
$n(z)$ to visualize the electron redistribution upon formation of the
interface
\begin{equation}
\label{eq2}
\Delta n (z) = n_{\rm M \vert G} (z) - n_{\rm M}(z) - n_{\rm G}(z),
\end{equation}
where $n_{\rm M \vert G}(z)$, $n_{\rm M}(z)$ and $n_{\rm G}(z)$ denote
the plane averaged densities of the graphene-covered metal, the clean
metal surface, and free-standing graphene, respectively. Notice that the
structure of the clean metal surface is required to be the same as that
of the graphene-covered metal surface.
The results for graphene physisorbed on Al, Cu, Ag, Au and Pt, are
shown in Fig.~\ref{ref:fig4b}. $\Delta n$ is localized near the
interface for all metal substrates and in the majority of cases it has
the shape of a simple dipolar charge distribution.

We estimate the charge $q$ (per carbon atom) that is responsible for
the dipole by integrating $\Delta n$ from the node at $z_0$ between the
metal surface and the graphene sheet,
\begin{equation}
\label{eq3}
q = e \, \int_{z_{0}}^{\infty}dz\, \Delta n (z)/N_{\rm C},
\end{equation}
where $N_{\rm C}$ is the number of
carbon atoms in the unit cell; $-e$ is the charge of an electron. These numbers are included in
Fig.~\ref{ref:fig4b}. The sign and size of the dipole charges are
consistent with the changes of the metal work function upon adsorption
of graphene. Note that relatively small values of charge transfer give rise to
quite substantial work function changes, see Table~\ref{ref:tab1}.

The above analysis points to the use of a plane capacitor model to
describe the potential step $\Delta V$. As sketched in
Fig.~\ref{ref:fig4a}, the charge distribution is then modelled as two
sheets of charge $\pm q$. Since the charge is predominantly localized
between graphene and the metal surface, the effective distance $z_d$
between the charge sheets should be smaller than the graphene
metal separation $d$.

If the interaction between graphene and the metal surface is weak, as
in the case of physisorption, one naively expects that electrons will
be transferred to graphene if the clean metal work function is lower
than that of free-standing graphene, i.e., if $W_{\rm M} < W_{\rm G}$.
Electrons should then flow from graphene to the metal surface if
$W_{\rm M}> W_{\rm G}$ and the crossover point from $n$- to $p$-type
doping would be exactly at $W_{\rm M}=W_{\rm G}$. The results obtained
for the equilibrium graphene-metal separation, $d \sim 3.3$ \AA\ in
Fig.~\ref{ref:fig3}, do not confirm this simple picture of the doping
mechanism. Instead, the crossover point between $n$- and $p$-type
doping is found for a metal with a work function $W_{\rm M}=W_{\rm G} +
0.9$ eV.

This simple picture of charge transfer cannot be entirely wrong. If the
graphene-metal separation is increased, the crossover point from $n$-
to $p$-type doping decreases to its expected value, $W_{\rm M}\sim
W_{\rm G}$, for large separations. This is illustrated by the upper
curve in Fig.~\ref{ref:fig3} calculated for a graphene-metal separation
of $d = 5.0$~\AA. It clearly indicates that, at the equilibrium
separation $d_{\rm eq} \sim 3.3$~\AA, the charge reordering at the
graphene-metal interface is the result not only of a charge transfer
between metal and graphene electronic levels that equilibrates the
graphene and metal Fermi energies but that there is also a contribution
from a direct interaction between the metal and graphene. A similar
interaction, which has a significant repulsive contribution, plays an
important role in describing the dipole formation for closed shell
atoms and organic molecules adsorbed upon metal
surfaces.\cite{Silva:prl03,Rusu:thesis07} The interaction depends on
the wave function overlap between the metal and the adsorbed species.
We expect it therefore to be very sensitive to the metal-graphene
separation $d$ and to vanish exponentially with increasing $d$.

\subsection{Phenomenological model}
\label{subsec:wmgbinding} In this section we construct a simple and
general model to describe the Fermi level and work function shifts
calculated from first-principles for graphene physisorbed on Al, Ag,
Cu, Au, and Pt substrates. All relevant parameters are shown in
Fig.~\ref{ref:fig4a}. We start by writing the work function of the
graphene-covered metal as $W(d) = W_{\rm M} - \Delta V(d)$, where
$\Delta V$ is the potential step generated by the interface dipole
layer. Its size depends on the graphene-metal separation $d$. The Fermi
level shift in graphene and the work function are related by Eq.
(\ref{eq1}). Use of these relations implicitly assumes that the
graphene electronic energy levels around the Fermi energy are
essentially unchanged by the interaction between graphene and the metal
and that the band structure of graphene is just rigidly shifted by the
interface potential $\Delta V$.

A key element of the model is to write the interface potential step as
$\Delta V(d) = \Delta_{\rm tr}(d) + \Delta_{\rm c}(d)$. The first term,
$\Delta_{\rm tr}(d)$, results from the direct charge transfer between
graphene and the metal, which is driven by the difference in work
functions. The second term, $\Delta_{\rm c}(d)$, describes the
short-range interaction discussed in the previous section, that results
from the overlap of the metal and graphene wave functions. We
parameterize it as
\begin{equation}
\label{eq:dc}
\Delta_{\rm c}(d)=e^{-\gamma d}\, (a_0 + a_1 d + a_2 d^2),
\end{equation}
i.e., we assume that it vanishes exponentially with increasing
graphene-metal separation $d$. The exact asymptotic functional
dependence of $\Delta_{\rm c}(d)$ for large $d$ would very likely
change if one were to go beyond DFT/LDA and take the van der Waals
interaction into account. The asymptotic form of $\Delta_{\rm c}(d)$
is, however, not important because $|\Delta_{\rm c}(d)|$ becomes
negligible for large $d$ anyway.

To model the electron transfer contribution, $\Delta_{\rm tr}(d)$, we
use a plane capacitor model so $\Delta_{\rm tr}(d)=\alpha N(d) z_d$,
where $\alpha = e^2/\varepsilon_0 A=34.93$ eV/\AA\, with $A = 5.18$
\AA$^2$ the area of the graphene unit cell, and $N(d)$ is the number of
electrons (per unit cell) transferred from graphene to the metal. Note
that $N(d)$ becomes negative if electrons are transferred from the
metal to graphene. The parameter $z_d$ is the effective distance
between the sheets of transferred charge on graphene and the metal. It
is smaller than the geometrical metal-graphene separation, $z_d < d$,
because most of the charge is localized between the graphene layer and
the metal surface. We approximate the effective distance between the
charge sheets by $z_d = d - d_0$ with $d_0$ a constant.

\begin{figure}[!tpb]
\begin{center}
\includegraphics[width=1.0\columnwidth]{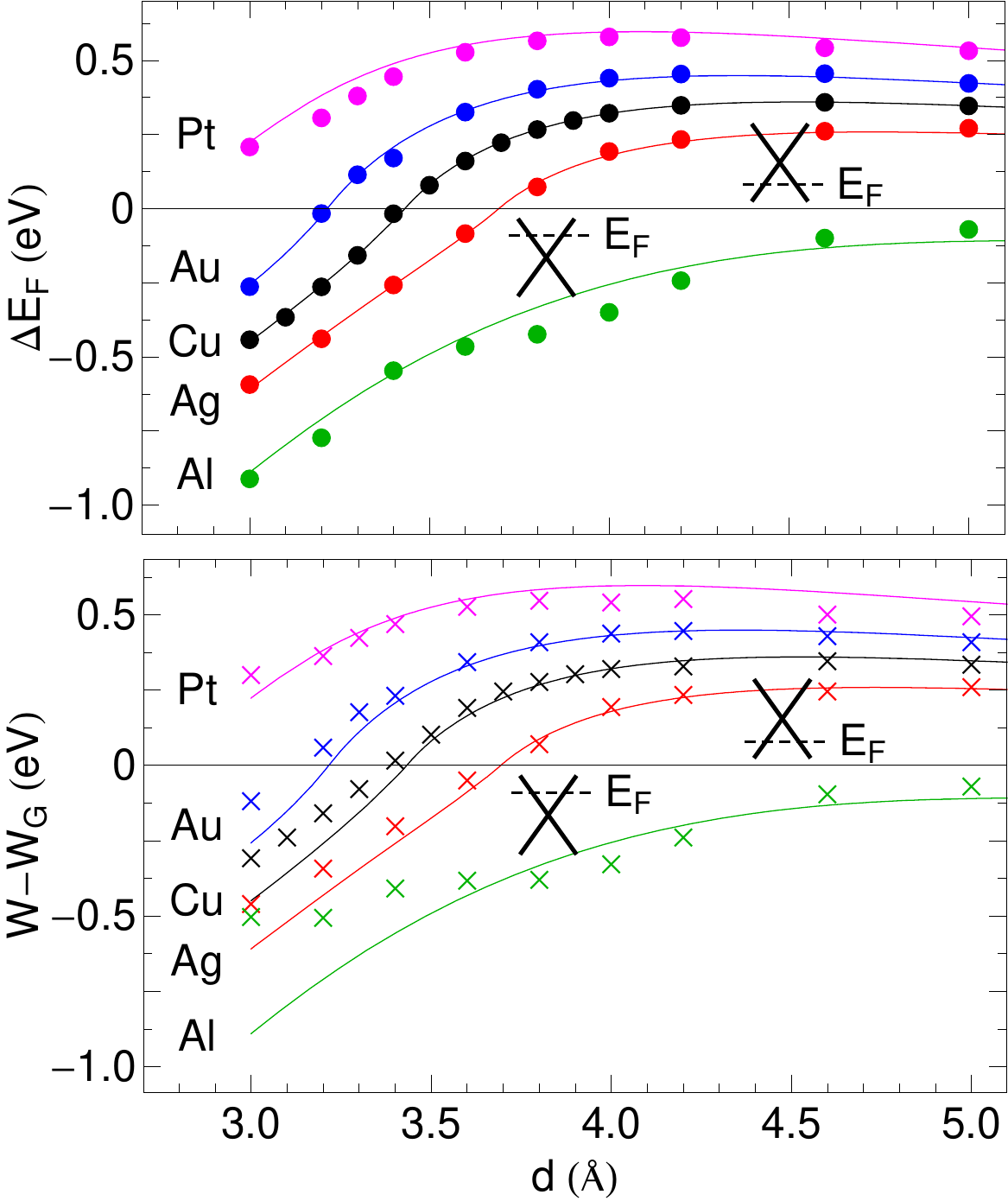}
\end{center}
\caption{(Color online) Top panel: Fermi level shifts relative to the
Dirac point, $\Delta E_{\rm F}(d)$, as a function of the graphene-metal
separation $d$ for physisorbed graphene. Bottom panel: calculated work
functions $W(d)$ relative to that of a free graphene sheet, $W_{\rm
G}$. The dots (top) and crosses (bottom) give the calculated DFT/LDA
results, the solid lines describe the results obtained from the
phenomenological model, Eq.~(\ref{eqn1}).\cite{footnote2}}
\label{ref:fig5}
\end{figure}

A closed set of equations for the Fermi level shift $\Delta E_{\rm F}$
and the work function $W$ is obtained by determining the relation
between $\Delta E_{\rm F}$ and the number of electrons $N$ transferred
between graphene and the metal. For an energy range within $\pm 1$~eV
of the conical points, the graphene density of states is described well
by a linear function
\begin{equation}
\label{eq4}
D(E)=D_0 |E|,
\end{equation}
with $D_0 = 0.09$/(${\rm eV}^2$ unit cell). Integrating the density of
states from the neutrality point,
$\int_0^{\Delta E_{\rm F}} dE\, D(E)$, yields the required relation,
$N = {\rm sign}(\Delta E_{\rm F}) D_0 \Delta E_{\rm F}^2/2$.
\cite{footnote1}

The model can be summarized by the set of equations
\begin{equation}
\label{eq:all}
\begin{cases}
W(d) = W_{\rm M} - \Delta V(d),\\
\Delta V(d) = \Delta_{\rm tr}(d) + \Delta_{\rm c}(d), \\
\Delta_{\rm tr}(d) = \alpha\, N(d)\, (d - d_0),  \\
N(d) = {\rm sign}(\Delta E_{\rm F})\, \frac{1}{2} D_0 \, \Delta E_{\rm F}(d)^2, \\
\Delta E_{\rm F}(d) = W(d) - W_{\rm G}.
\end{cases}
\end{equation}
Solving this set results in the following simple expression for the
Fermi level shift
\begin{equation}
\Delta E_{\rm F} (d)\hspace{-0.5mm} = \hspace{-0.5mm} \pm
\frac{ \sqrt{1\hspace{-0.5mm} +\hspace{-0.5mm} 2 \alpha D_0 ( d \hspace{-0.5mm} - \hspace{-0.5mm} d_0 ) \vert  W_{\rm M}\hspace{-0.5mm} -\hspace{-0.5mm} W_{\rm G}\hspace{-0.5mm} -\hspace{-0.5mm} \Delta_{\rm c}(d) \vert } \hspace{-0.5mm}- 1 }{ \alpha D_0  ( d\hspace{-0.5mm} -\hspace{-0.5mm} d_0 )},
\label{eqn1}
\end{equation}
where the sign of $\Delta E_{\rm F}$ is determined by the sign of
$W_{\rm M}-W_{\rm G}-\Delta_{\rm c}$. The work function of the
graphene-covered metal surface is then obtained from Eq. (\ref{eq1}).

We used first-principles calculations to obtain $\Delta E_{\rm F}$ and
$W$ explicitly for a range of separations $d$. These first-principles
data points are shown in Fig.~\ref{ref:fig5}. We could now fit the
model to these points and so obtain a simple interpretation of the
numerical results. However, we can do better than that. It turns out
that the parameter $d_0$ and the function $\Delta_{\rm c}(d)$ depend
only very weakly on the metal substrate. This means that we can fit
these quantities to the first-principles results for a single metal
substrate and subsequently use $\Delta_{\rm c}(d)$ and $d_0$ as
universal parameters to predict the Fermi level shifts in graphene for
all metal substrates. To determine $d_0$ and $\Delta_{\rm c}(d)$ we use
the DFT results for graphene on the Cu(111) surface. \cite{footnote2}

The Fermi level shift for graphene physisorbed on any other metal
substrate can then be obtained from the model, using only the work
function of the clean metal surface, $W_{\rm M}$, and that of
free-standing graphene, $W_{\rm G}$, as input parameters in
Eq.~(\ref{eqn1}). The accuracy of this model is demonstrated by
Fig.~\ref{ref:fig3} and the top panel of Fig.~\ref{ref:fig5}.
 The latter shows that the distance dependence of the Fermi
level shift is represented very well by the model. For graphene on Cu,
Ag, and Au(111) the deviations from the first-principles results are
$\ll 0.1$ eV, whereas for graphene on Al and Pt(111) they are still
$\lesssim 0.1$ eV. The slightly larger deviations for Al and Pt might
be due to a more complex interaction between graphene and these
surfaces, caused by the open $p$ and $d$ shells of Al and Pt,
respectively. The latter might also be responsible for a non-monotonic
behavior of $\Delta E_{\rm F}(d)$ and $W(d)$ for Al and Pt in Fig.~\ref{ref:fig5}.

Once $\Delta E_{\rm F}$ has been determined, the work function $W$ of
the metal-graphene system can be obtained using Eq.~(\ref{eq1}) and the
results are shown in the bottom panel of Fig.~\ref{ref:fig5}. The work
function given by the model agrees with the first-principles results
within $\lesssim 0.2$ eV for $d\gtrsim 3.3$ \AA, the equilibrium
separation, but seems to be systematically lower for smaller values of
$d$. The difference becomes smaller upon increasing $d$. For small
graphene metal separations $d$, Eq.~(\ref{eq1}) no longer holds
exactly, as the electronic structure of graphene is perturbed by the
presence of the metal substrate.
Note, however, that 0.2 eV is comparable to the difference between the
experimental and the DFT/LDA work functions, see Table \ref{ref:tab1},
which means that the error resulting from the model is tolerable.

According to Eq.~(\ref{eq:all}) one can calculate the sign and
concentration of the charge carriers in graphene, $N$, from the Fermi
level shift $\Delta E_{\rm F}$. The {\em nominal} charge on graphene
per unit cell containing two carbon atoms, $eN$, is $-14.6$, $-1.6$,
$-1.3$, $1.6$, $4.9\times 10^{-3}\,e$ for graphene on Al, Ag, Cu, Au,
Pt(111), respectively.\cite{footnote1}  Although these charges are very
small, the Fermi level shifts are quite substantial because the
graphene DOS close to the conical points is so low, but they are still
within the linear regime described by Eq.~(\ref{eq4}). In the
terminology used in semiconductor physics this amount of charge per
unit cell would be classified as heavy doping.

The crossover from $p$- to $n$-type doping occurs when the Fermi level
coincides exactly with the conical points of graphene, i.e., $\Delta
E_{\rm F}=0$. According to Eq.~(\ref{eqn1}), this happens if the work
function of the metal is given by the critical value
\begin{equation}
W_0(d) = W_{\rm G} + \Delta_{\rm c}(d).
\end{equation}
The critical work function $W_0$ depends on the graphene-metal
separation $d$, since the term $\Delta_{\rm c}$ resulting from the
direct graphene-metal interaction depends strongly on $d$, becoming
negligible if $d\gtrsim 4.2$ \AA. $W_0$ then approaches $W_{\rm G} =
4.5$ eV, i.e., the critical work function is that of free-standing
graphene. However, at the equilibrium graphene metal separation,
$d_{\rm eq} = 3.3$ \AA, $\Delta_{\rm c}\approx 0.9$ eV, leading to a
critical work function $W_{0}(d)\approx 5.4$ eV. Though the short-range
graphene-metal interaction does not significantly change the graphene
band structure, it does lead to a sizeable potential step at the
equilibrium separation which is downwards from metal to graphene as
indicated in Fig.~\ref{ref:fig4a}. The size of this potential step is
relatively insensitive to the metal substrate. A similar potential step
has been observed in the adsorption of closed-shell molecules on metal
surfaces, where it has been interpreted in terms of an exchange
repulsion between the electrons on the molecules and the metal
substrate.\cite{Rusu:thesis07}

The phenomenological model we have outlined describes the doping of
graphene by metal contacts and the work function shifts caused by
adsorption of graphene when the graphene-metal bonding is weak. The
model is based on the linearity of the graphene DOS, which holds for
energies within $\pm 1$ eV around the conical points. Therefore, the
criterion for the validity of the model is $\Delta_{\rm g}\ll
\vert\Delta E_{\rm F}\vert\lesssim 1$ eV, where $\Delta_{\rm g}$ is a
band gap induced in graphene by interaction with the substrate.
\cite{Giovannetti:prb07}

\subsection{Chemisorbed graphene}
\label{subsec:smgbinding}

In the previous sections we defined the doping of physisorbed graphene
in terms of the Fermi level shift $\Delta E_{\rm F}$ with respect to
the conical points in the graphene band structure. Negative and
positive $\Delta E_{\rm F}$ correspond, respectively, to $n$-type and
$p$-type doping. This procedure cannot be used for graphene that is
chemisorbed on the Ni, Co, or Pd(111) or on the Ti(0001) surface
because the strong graphene-metal bonding interaction destroys the
conical points; see the Ni, Co and Pd panels in Fig.~\ref{ref:fig2}.

\begin{figure}[!tpb]
\includegraphics[width=1.0\columnwidth]{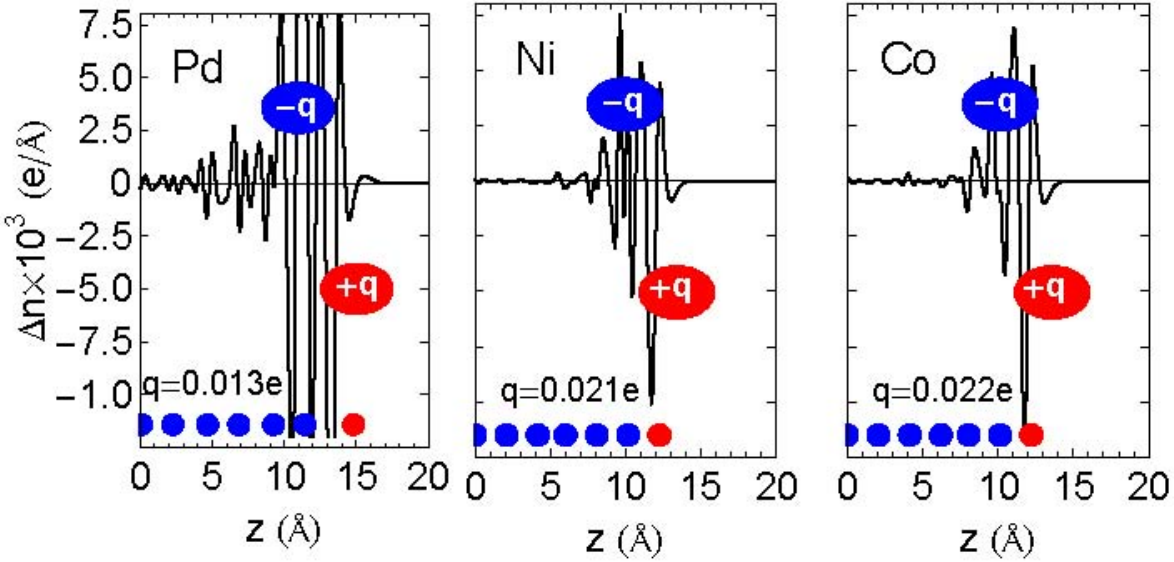}
\caption{(Color online) Plane-averaged electron difference density
$\Delta n (z)$ (per unit cell) showing the charge displacement upon
formation of the chemisorbed graphene-M(111) interface, with M = Ni,
Co, Pd. $q$ is the charge (per carbon atom) calculated by integrating
$\Delta n (z)$ from the central nodal point to
infinity.\cite{footnote1}} \label{ref:fig4c}
\end{figure}

The more complex bonding of chemisorbed graphene is also illustrated by
comparing the plane-averaged electron difference densities $\Delta
n(z)$ shown in Figs.~\ref{ref:fig4b} and \ref{ref:fig4c}. In the
physisorption case, $\Delta n$ still has the characteristics of a
simple interface dipole while in the chemisorption case it is much more
complicated, indicating the formation of new bonds at the interface
between graphene and Ni, Co and Pd. The charge reordering at the
interface upon chemisorption is substantial, as reflected by the larger
values for $q$ given in Fig.~\ref{ref:fig4c}. This leads to
considerable shifts in the metal work functions upon chemisorption of
graphene. In all cases studied in this paper graphene acts as an
electron donor, lowering the metal work function. For Ni, Co and
Pd(111) the work function lowering is 1.81, 1.66, and 1.64 eV,
respectively, and for Ti(0001) it is 0.56 eV.

\begin{figure}[!tpb]
\begin{center}
\includegraphics[width=0.9\columnwidth]{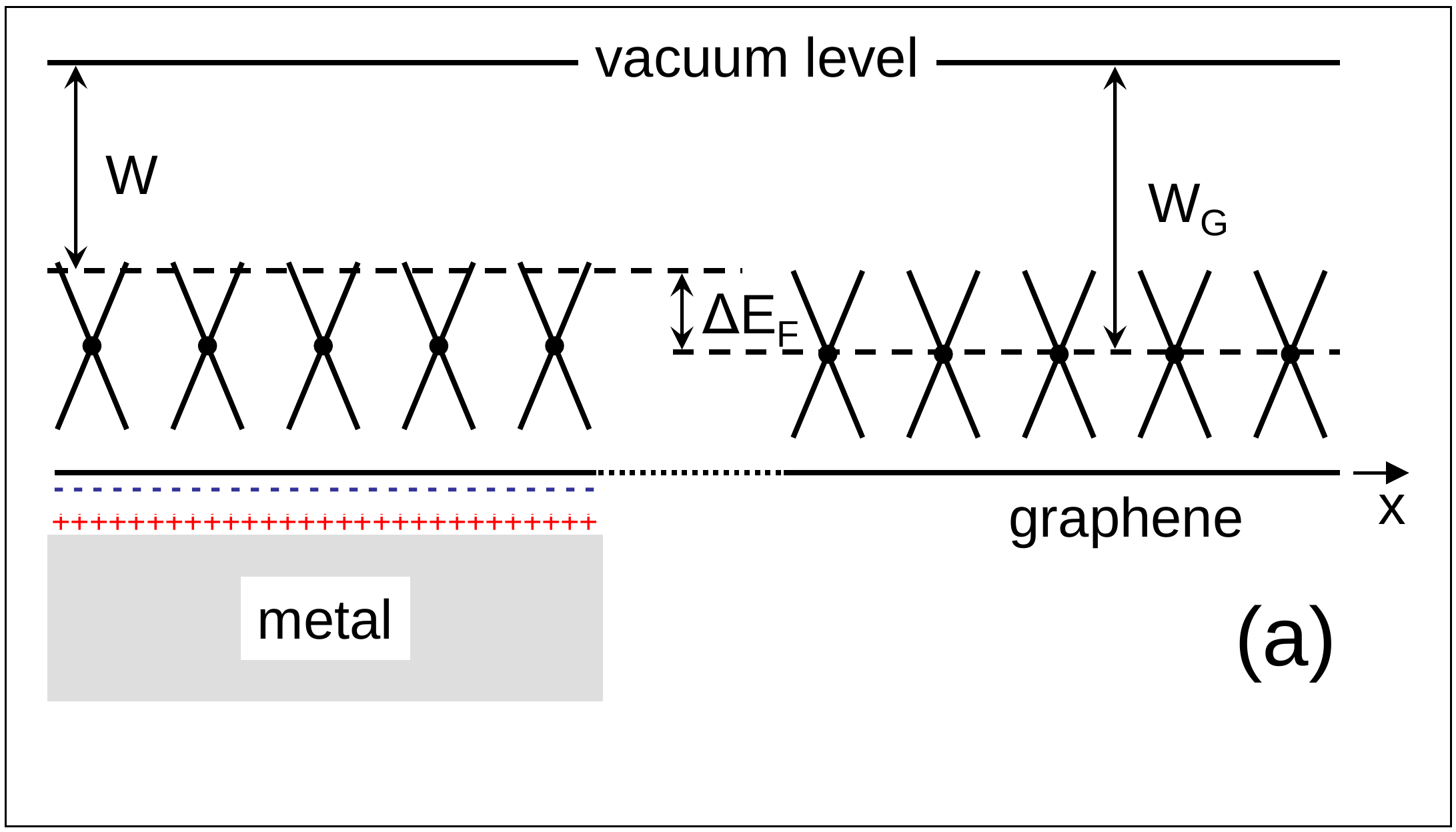}
\includegraphics[width=0.9\columnwidth]{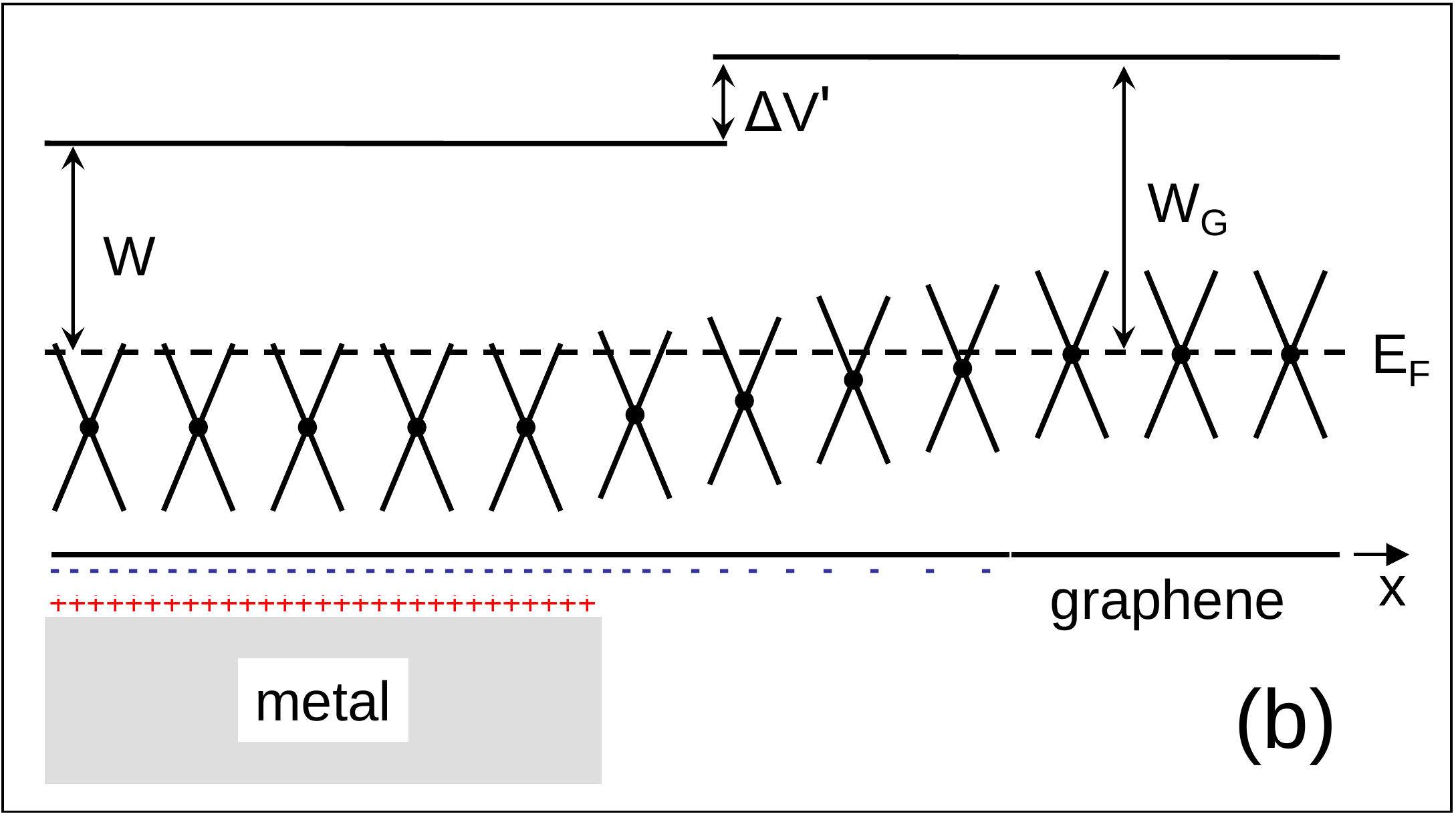}
\end{center}
\caption{(Color online) Schematic representation of a graphene sheet
partly in contact with a metal electrode. The work function of the
graphene-covered metal electrode, $W$, is here smaller than the work
function of graphene, $W_{\rm G}$, and the graphene sheet becomes
$n$-type doped. Far from the electrode the conical point of graphene
(bold dots) approaches the Fermi energy. (a) In the absence of
communication between the graphene covered metal and the intrinsic
graphene sheet, there is a discontinuity $\Delta E_{\rm F}$ of the
Fermi energies. (b) As soon as the two systems can communicate,
equilibration of the Fermi energies takes place by the transfer of
electrons from the low to the high work function system and the joint
Fermi energy is fixed by the graphene-covered metal electrode, $E_{\rm
F}=-W$. This rearrangement of charge gives rise to a potential shift
$\Delta V^{\prime}$. The
band-bending and graphene doping depend on the distance $x$ from the
contact.} \label{ref:fig6}
\end{figure}

Despite the impossibility of identifying a Fermi level shift and
therefore the type and magnitude of the doping from a simple
examination of the band structures in chemisorbed graphene, it is
possible to define a suitable measure for application to
current-in-plane (CIP) geometries. In a CIP geometry only part of the
graphene sheet is covered by metal electrodes, and the greater part of
the sheet is ``free-standing''. The type and effective concentration of
charge carriers in graphene contacted to metallic leads can be measured
in experiments using the CIP geometry shown schematically in
Fig.~\ref{ref:fig6} (for clarity, for the physisorbed case). At a large
distance from the metal contact the Fermi level in free-standing
graphene approaches the conical points. At the metallic contact the
Fermi level is fixed by the interaction with the metal electrode. The
difference between the Fermi levels in the adsorbed and free-standing
graphene is given by the difference between the work function of the
graphene-covered metal surface, $W$, and that of free-standing
graphene, $W_{\rm G}$, or in other words, by Eq. (\ref{eq1}); see
Fig.~\ref{ref:fig6}(a). We already used this relation in our
description of physisorbed graphene. Since the work function $W$ can be
determined for chemisorbed as well as physisorbed graphene, it can be
applied to all metal electrodes in the CIP geometry.

To accommodate the Fermi level difference, charge transfer takes place
between the contacted and free-standing regions.\cite{footnote3}
Alignment of the Fermi levels results in the band bending shown
schematically in Fig.~\ref{ref:fig6}(b). The amount of band bending is
given by Eq.~\eqref{eq1}. The band bending region is $p$- or $n$-type
doped, depending on the sign of $\Delta E_{\rm F}$. If $\Delta E_{\rm
F}>0$, graphene is $p$-type doped, and if $\Delta E_{\rm F}<0$, it is
$n$-type doped. Different metal electrodes can then be used to make
$p$-$n$ junctions in graphene.

For graphene chemisorbed on Ni, Co, Pd(111) and Ti(0001) surfaces, we
find $\Delta E_{\rm F}\equiv W-W_{\rm G}=-0.82$, $-0.70$, $-0.45$, and
$-0.34$ eV, respectively, see Table~\ref{ref:tab1}. Chemisorption of
graphene on these surfaces lowers their work function to below that of
free standing graphene. Therefore, we expect graphene to be $n$-doped
by these metal electrodes.

\subsection{Sensitivity to approximations}
\label{sec:functional} The calculations on the interaction between
graphene and the metal substrates we have discussed so far, are at the
level of DFT/L(S)DA. The results given in Table~\ref{ref:tab1} show
that calculated work functions of clean metal surfaces agree with
experimental data within $\sim 0.2$ eV. A similar agreement is observed
between calculated work functions of graphene-covered metal surfaces
and available experimental data. The change in work function upon
graphene adsorption is determined by the formation of an interface
dipole and the charge transfer between metal and graphene. Apparently
this charge redistribution at the interface is described rather well by
the LDA functional.

\begin{figure}[!tpb]
\begin{center}
\includegraphics[width=1.0\columnwidth]{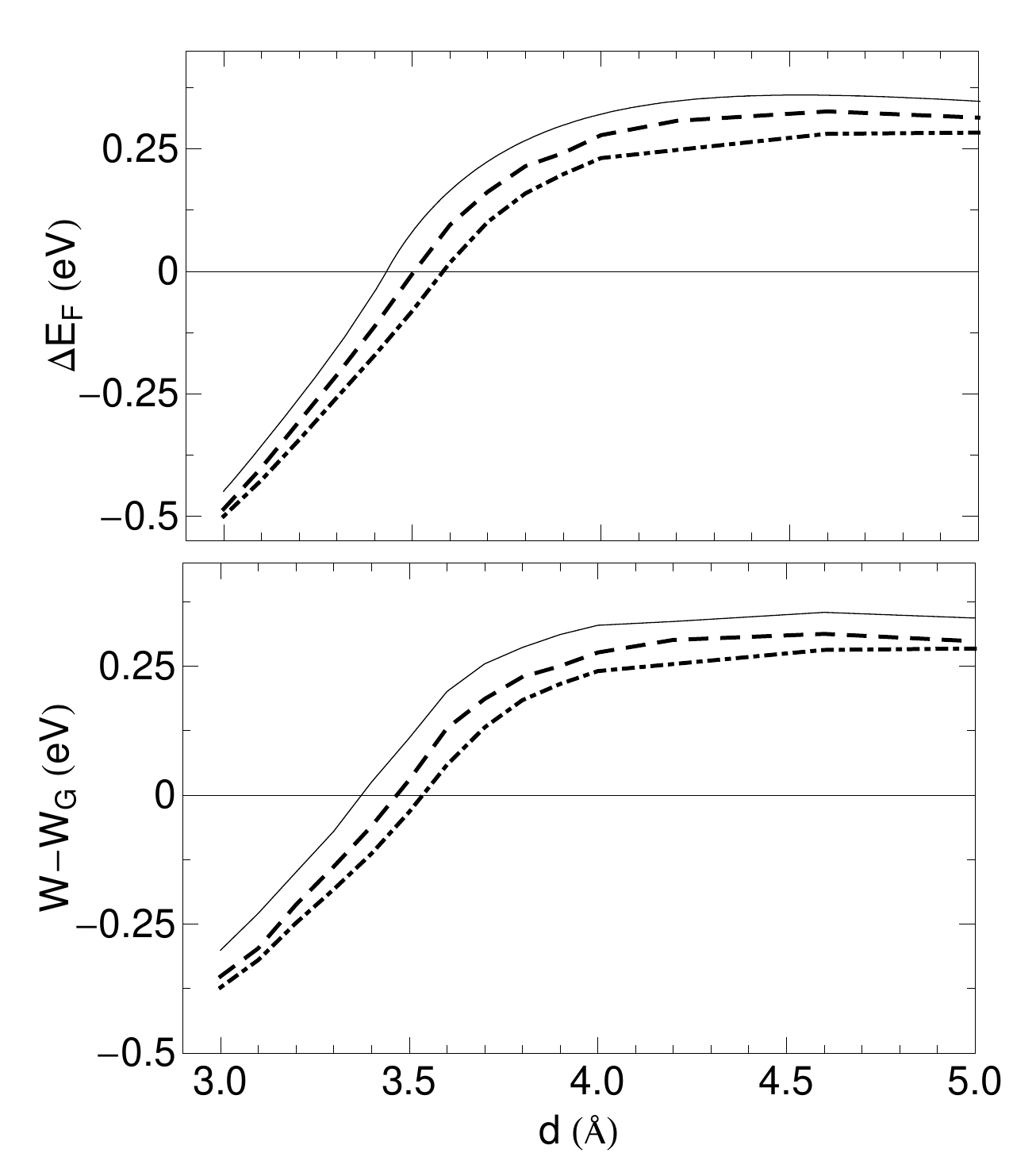}
\end{center}
\caption{(Color online) The dashed lines correspond to the
interpolation curves for $\Delta E_{\rm F}$ and $W(d) - W_{\rm G}$ as a
function of the graphene-Cu(111) surface distance $d$ calculated with
the PW91 GGA functional. The dash-dotted lines represent the results
obtained with the LDA functional with graphene stretched to the Cu(111)
lattice constant $a_{\rm hex}=2.49$\,\AA. As a reference, the solid
lines give the LDA results obtained with the optimized graphene lattice
constant $a_{\rm hex}=2.445$\,\AA.} \label{ref:fig7}
\end{figure}

\textit{Semi-local GGA functionals} are frequently preferred in DFT
calculations. For graphene that is physisorbed on a metal surface the
commonly used GGA functionals give an interaction that is either too
weak or even purely repulsive making it impossible to predict the
equilibrium distance between graphene and the metal surface. In order
to test whether the Fermi level shift in adsorbed graphene and its work
function are sensitive to the particular choice of the density
functional, we have calculated these parameters for graphene on Cu(111)
as a function of the graphene-metal surface distance, using the PW91
GGA functional.\cite{Perdew:prb92,Perdew:prb93} The results are shown
in Fig.~\ref{ref:fig7}. The Fermi level shifts $\Delta E_{\rm F}$
calculated with GGA are within $\sim 0.07$ eV of the ones obtained with
LDA. The same holds for the work function difference $W-W_{\rm G}$,
which means that the description of the doping of graphene is not
sensitive to the functional used in the calculations. However, the
absolute values of $W$ and $W_{\rm G}$ calculated using the GGA differ
from those obtained using the LDA by $\sim 0.1-0.3$ eV, which is a
typical difference between work functions obtained with these two
density functionals.\cite{Rusu:thesis07}

Another issue is the possible effect of a truly \textit{non-local van
der Waals interaction} between the graphene sheet and metal surface,
which is neglected in our study. Van der Waals interactions have been
addressed recently in calculations on graphite, hexagonal boron-nitride
and diatomic molecules of inert
gases.\cite{Rydberg:prl03,Thonhauser:prb07} A non-local correction to a
GGA-type density functional has been proposed in
Ref.~\onlinecite{Thonhauser:prb07} that correctly reproduces the
asymptotic van der Waals tail of the binding energy curve at large
intermolecular distances. This non-local correction has little impact
on the charge distribution, however.\cite{Thonhauser:prb07} Since it is
the charge distribution that gives the Fermi level shift, the work
function and the doping of adsorbed graphene, we conclude that these
quantities are adequately described by local or semi-local functionals.

In our calculations we choose the in-plane lattice constant of graphene
equal to its optimized LDA value $a = 2.445$ \AA, and adapt the lattice
constants of the metals accordingly. The approximation made by this
matching procedure seems reasonable, since the mismatch with the
lattice parameters of the metal (111) surfaces is only 0.8-3.8\%, see
Table~\ref{ref:tab1}. The \textit{largest lattice mismatch} is that
between graphene and Cu(111). To estimate the error on the Fermi level
shift and the work function of adsorbed graphene, we have calculated
these quantities while stretching graphene to the LDA optimized
in-plane lattice constant of Cu(111) (2.49 \AA). The results are shown
in Fig.~\ref{ref:fig7}. The Fermi level shifts and work function
differences are within 0.15 eV of the results obtained with the
optimized graphene lattice constant. One expects to see a smaller
effect for other metals, as the lattice mismatch between graphene and
other metals is smaller.

\section{Discussion and conclusions}
\label{sec:conclusions} The theoretical study performed in the previous
sections assumes that the graphene sheet is adsorbed on a clean
crystalline metal contact. We have used the model represented by
Eq.~(\ref{eqn1}) to describe the Fermi level shift in physisorbed
graphene. Interpreting experiments that are not carried out in
ultra-high vacuum requires some modifications because of impurities
that will be present at the metal-graphene interface.\cite{Lee:natn08}
The work function of the clean metal surface must then be replaced with
that of the metal surface contaminated by water molecules, oxygen
and/or nitrogen, for instance. The short-range graphene-metal surface
interaction, represented by the potential term $\Delta_{\rm c}$, and
the equilibrium separation $d_{\rm eq}$ will also be affected by such
adsorbates and the effects will depend critically upon the adsorbate
concentration. The same obviously holds for chemisorbed graphene. A
large concentration of adsorbates on the metal surface could break the
chemical interaction between graphene and the metal.

In some experiments the interface between the graphene sheet and metal
surface might contain a thicker buffer layer consisting, for example,
of water, or a metal-oxide.\cite{Lee:natn08} That would certainly
modify the graphene-substrate interaction. An input parameter to the
model of Eq.~(\ref{eqn1}) then is the work function of the metal with
the buffer layer on top. Using the plane capacitor model one should
replace $\alpha$ in Eq.~(\ref{eqn1}) by $\alpha/\kappa$, where $\kappa$
is the effective dielectric constant of the buffer layer. Obviously the
distance $z_d$ between the charge sheets has to be modified accordingly
to take the thickness of the buffer layer into account. In addition the
potential term $\Delta_{\rm c}$, which represents the short-range
interaction of graphene with the substrate should now reflect the
graphene-buffer layer interaction.

A metallic buffer layer consisting of, for instance, Ti is often used
in experiments to establish a good contact between graphene and the
electrodes. Such layers have a thickness of typically $\sim 5$ nm,
which is sufficiently thick that the contact should be considered as a
graphene-Ti contact.

Using first-principles DFT calculations, we have systematically studied
the interaction and charge transfer between graphene and a range of
metal surfaces with different work functions. We found that graphene is
chemisorbed on Co, Ni, and Pd(111) surfaces, and on the Ti(0001)
surface and that this strong interaction perturbs the electronic
structure of graphene significantly. In contrast, adsorption of
graphene on Al, Cu, Ag, Au and Pt(111) surfaces leads to weak bonding,
which preserves the typical graphene electronic structure, including
the conical points. Even in this physisorbed case, however, there is a
short-range graphene-metal interaction, as well as a charge transfer
between the graphene and metal states. These result in a doping of
graphene, i.e., a shift of the Fermi level with respect to the conical
points, and a significant change in the work function of the
graphene-covered metal surfaces, as compared to the clean metal
surface.

To extend the applicability of the DFT results on physisorbed graphene
we develop a simple general model, which takes into account the charge
transfer between the graphene and metal states and the short-range
graphene-metal interaction. We find that the latter only weakly depends
on the metal. Therefore it can be fitted using the DFT results on one
metal substrate. The model then only has the work functions of
free-standing graphene and that of the clean metal surface as input
parameters and it predicts the Fermi level shift and carrier
concentration in graphene, as well as the work function of the
graphene-covered metal substrate. We find that graphene is $n$-type
doped if the metal work function $W_{\rm M} \lesssim 5.4$ eV, whereas
it is $p$-type doped if $W_{\rm M} \gtrsim 5.4$ eV.

For the CIP geometry, where only part of the graphene sheet covers the
metal electrode, we propose a definition of the doping that is based
upon the work function of graphene-covered metal surface. This
definition is valid both for the chemisorbed and physisorbed cases. It
predicts that adsorption of graphene on Al, Ag, Cu, Co, Ni, Pd(111)
surfaces and on the Ti(0001) surface leads to $n$-type doping. The high
values of the work functions of Au and Pt(111) substrates lead to
$p$-type doping of graphene.

Both the analytical model and the CIP definition of the graphene doping
are derived from general principles and should be applicable to any
metal surface on which graphene can be epitaxially grown. This opens up
the possibility of a general understanding of $p$-$n$ junctions
prepared by doping graphene with metal contacts.\cite{Khomyakov:unpub2}

\acknowledgments This work is part of the research programs of
``Chemische Wetenschappen (CW)'' and the ``Stichting voor Fundamenteel
Onderzoek der Materie (FOM)'', both financially supported by the
``Nederlandse Organisatie voor Wetenschappelijk Onderzoek (NWO)'', and
of ``NanoNed'', a nanotechnology program of the Dutch Ministry of
Economic Affairs. Part of the calculations were performed with a grant
of computer time from the ``Stichting Nationale Computerfaciliteiten
(NCF)''.


\begin{thebibliography}{66}
\expandafter\ifx\csname natexlab\endcsname\relax\def\natexlab#1{#1}\fi
\expandafter\ifx\csname bibnamefont\endcsname\relax
  \def\bibnamefont#1{#1}\fi
\expandafter\ifx\csname bibfnamefont\endcsname\relax
  \def\bibfnamefont#1{#1}\fi
\expandafter\ifx\csname citenamefont\endcsname\relax
  \def\citenamefont#1{#1}\fi
\expandafter\ifx\csname url\endcsname\relax
  \def\url#1{\texttt{#1}}\fi
\expandafter\ifx\csname urlprefix\endcsname\relax\def\urlprefix{URL }\fi
\providecommand{\bibinfo}[2]{#2}
\providecommand{\eprint}[2][]{\url{#2}}

\bibitem[{\citenamefont{Kroto et~al.}(1985)\citenamefont{Kroto, Heath,
  {O'Brien}, Curl, and Smalley}}]{Kroto:nat85}
\bibinfo{author}{\bibfnamefont{H.~W.} \bibnamefont{Kroto}},
  \bibinfo{author}{\bibfnamefont{J.~R.} \bibnamefont{Heath}},
  \bibinfo{author}{\bibfnamefont{S.~C.} \bibnamefont{{O'Brien}}},
  \bibinfo{author}{\bibfnamefont{R.~F.} \bibnamefont{Curl}}, \bibnamefont{and}
  \bibinfo{author}{\bibfnamefont{R.~E.} \bibnamefont{Smalley}},
  \bibinfo{journal}{Nature} \textbf{\bibinfo{volume}{318}},
  \bibinfo{pages}{162} (\bibinfo{year}{1985}).

\bibitem[{\citenamefont{Iijima}(1991)}]{Iijima:nat91}
\bibinfo{author}{\bibfnamefont{S.}~\bibnamefont{Iijima}},
  \bibinfo{journal}{Nature} \textbf{\bibinfo{volume}{354}}, \bibinfo{pages}{56}
  (\bibinfo{year}{1991}).

\bibitem[{\citenamefont{Novoselov
  et~al.}(2005{\natexlab{a}})\citenamefont{Novoselov, Jiang, Schedin, Booth,
  Khotkevich, Morozov, and Geim}}]{Novoselov:pnas05}
\bibinfo{author}{\bibfnamefont{K.~S.} \bibnamefont{Novoselov}},
  \bibinfo{author}{\bibfnamefont{D.}~\bibnamefont{Jiang}},
  \bibinfo{author}{\bibfnamefont{F.}~\bibnamefont{Schedin}},
  \bibinfo{author}{\bibfnamefont{T.~J.} \bibnamefont{Booth}},
  \bibinfo{author}{\bibfnamefont{V.~V.} \bibnamefont{Khotkevich}},
  \bibinfo{author}{\bibfnamefont{S.~V.} \bibnamefont{Morozov}},
  \bibnamefont{and} \bibinfo{author}{\bibfnamefont{A.~K.} \bibnamefont{Geim}},
  \bibinfo{journal}{Proc. Natl. Acad. Sci. U.S.A.}
  \textbf{\bibinfo{volume}{102}}, \bibinfo{pages}{10451}
  (\bibinfo{year}{2005}{\natexlab{a}}).

\bibitem[{\citenamefont{Novoselov et~al.}(2004)\citenamefont{Novoselov, Geim,
  Morozov, Jiang, Zhang, Dubonos, Grigorieva, and Firsov}}]{Novoselov:sc04}
\bibinfo{author}{\bibfnamefont{K.~S.} \bibnamefont{Novoselov}},
  \bibinfo{author}{\bibfnamefont{A.~K.} \bibnamefont{Geim}},
  \bibinfo{author}{\bibfnamefont{S.~V.} \bibnamefont{Morozov}},
  \bibinfo{author}{\bibfnamefont{D.}~\bibnamefont{Jiang}},
  \bibinfo{author}{\bibfnamefont{Y.}~\bibnamefont{Zhang}},
  \bibinfo{author}{\bibfnamefont{S.~V.} \bibnamefont{Dubonos}},
  \bibinfo{author}{\bibfnamefont{I.~V.} \bibnamefont{Grigorieva}},
  \bibnamefont{and} \bibinfo{author}{\bibfnamefont{A.~A.}
  \bibnamefont{Firsov}}, \bibinfo{journal}{Science}
  \textbf{\bibinfo{volume}{306}}, \bibinfo{pages}{666} (\bibinfo{year}{2004}).

\bibitem[{\citenamefont{Novoselov
  et~al.}(2005{\natexlab{b}})\citenamefont{Novoselov, Geim, Morozov, Jiang,
  Katsnelson, Grigorieva, Dubonos, and Firsov}}]{Novoselov:nat05}
\bibinfo{author}{\bibfnamefont{K.~S.} \bibnamefont{Novoselov}},
  \bibinfo{author}{\bibfnamefont{A.~K.} \bibnamefont{Geim}},
  \bibinfo{author}{\bibfnamefont{S.~V.} \bibnamefont{Morozov}},
  \bibinfo{author}{\bibfnamefont{D.}~\bibnamefont{Jiang}},
  \bibinfo{author}{\bibfnamefont{M.~I.} \bibnamefont{Katsnelson}},
  \bibinfo{author}{\bibfnamefont{I.~V.} \bibnamefont{Grigorieva}},
  \bibinfo{author}{\bibfnamefont{S.~V.} \bibnamefont{Dubonos}},
  \bibnamefont{and} \bibinfo{author}{\bibfnamefont{A.~A.}
  \bibnamefont{Firsov}}, \bibinfo{journal}{Nature}
  \textbf{\bibinfo{volume}{438}}, \bibinfo{pages}{197}
  (\bibinfo{year}{2005}{\natexlab{b}}).

\bibitem[{\citenamefont{Zhang et~al.}(2005)\citenamefont{Zhang, Tan, Stormer,
  and Kim}}]{Zhang:nat05}
\bibinfo{author}{\bibfnamefont{Y.~B.} \bibnamefont{Zhang}},
  \bibinfo{author}{\bibfnamefont{Y.~W.} \bibnamefont{Tan}},
  \bibinfo{author}{\bibfnamefont{H.~L.} \bibnamefont{Stormer}},
  \bibnamefont{and} \bibinfo{author}{\bibfnamefont{P.}~\bibnamefont{Kim}},
  \bibinfo{journal}{Nature} \textbf{\bibinfo{volume}{438}},
  \bibinfo{pages}{201} (\bibinfo{year}{2005}).

\bibitem[{\citenamefont{Zhou et~al.}(2006)\citenamefont{Zhou, Gweon, Graf,
  Fedorov, Spataru, Diehl, Kopelevich, Lee, Louie, and Lanzara}}]{Zhou:natp06}
\bibinfo{author}{\bibfnamefont{S.~Y.} \bibnamefont{Zhou}},
  \bibinfo{author}{\bibfnamefont{G.~H.} \bibnamefont{Gweon}},
  \bibinfo{author}{\bibfnamefont{J.}~\bibnamefont{Graf}},
  \bibinfo{author}{\bibfnamefont{A.~V.} \bibnamefont{Fedorov}},
  \bibinfo{author}{\bibfnamefont{C.~D.} \bibnamefont{Spataru}},
  \bibinfo{author}{\bibfnamefont{R.~D.} \bibnamefont{Diehl}},
  \bibinfo{author}{\bibfnamefont{Y.}~\bibnamefont{Kopelevich}},
  \bibinfo{author}{\bibfnamefont{D.~H.} \bibnamefont{Lee}},
  \bibinfo{author}{\bibfnamefont{S.~G.} \bibnamefont{Louie}}, \bibnamefont{and}
  \bibinfo{author}{\bibfnamefont{A.}~\bibnamefont{Lanzara}},
  \bibinfo{journal}{Nature Physics} \textbf{\bibinfo{volume}{2}},
  \bibinfo{pages}{595} (\bibinfo{year}{2006}).

\bibitem[{\citenamefont{Bolotin et~al.}(2008)\citenamefont{Bolotin, Sikes,
  Jiang, Fudenberg, Hone, Kim, and Stormer}}]{Bolotin:ssc08}
\bibinfo{author}{\bibfnamefont{K.~I.} \bibnamefont{Bolotin}},
  \bibinfo{author}{\bibfnamefont{K.~J.} \bibnamefont{Sikes}},
  \bibinfo{author}{\bibfnamefont{Z.}~\bibnamefont{Jiang}},
  \bibinfo{author}{\bibfnamefont{G.}~\bibnamefont{Fudenberg}},
  \bibinfo{author}{\bibfnamefont{J.}~\bibnamefont{Hone}},
  \bibinfo{author}{\bibfnamefont{P.}~\bibnamefont{Kim}}, \bibnamefont{and}
  \bibinfo{author}{\bibfnamefont{H.~L.} \bibnamefont{Stormer}},
  \bibinfo{journal}{Sol. State Comm.} \textbf{\bibinfo{volume}{146}},
  \bibinfo{pages}{351} (\bibinfo{year}{2008}).

\bibitem[{\citenamefont{Danneau et~al.}(2008)\citenamefont{Danneau, Wu,
  Craciun, Russo, Tomi, Salmilehto, Morpurgo, and Hakonen}}]{Danneau:prl08}
\bibinfo{author}{\bibfnamefont{R.}~\bibnamefont{Danneau}},
  \bibinfo{author}{\bibfnamefont{F.}~\bibnamefont{Wu}},
  \bibinfo{author}{\bibfnamefont{M.~F.} \bibnamefont{Craciun}},
  \bibinfo{author}{\bibfnamefont{S.}~\bibnamefont{Russo}},
  \bibinfo{author}{\bibfnamefont{M.~Y.} \bibnamefont{Tomi}},
  \bibinfo{author}{\bibfnamefont{J.}~\bibnamefont{Salmilehto}},
  \bibinfo{author}{\bibfnamefont{A.~F.} \bibnamefont{Morpurgo}},
  \bibnamefont{and} \bibinfo{author}{\bibfnamefont{P.~J.}
  \bibnamefont{Hakonen}}, \bibinfo{journal}{Phys. Rev. Lett.}
  \textbf{\bibinfo{volume}{100}}, \bibinfo{pages}{196802}
  (\bibinfo{year}{2008}).

\bibitem[{\citenamefont{Shon and Ando}(1998)}]{Shon:jpsj98}
\bibinfo{author}{\bibfnamefont{N.~H.} \bibnamefont{Shon}} \bibnamefont{and}
  \bibinfo{author}{\bibfnamefont{T.}~\bibnamefont{Ando}}, \bibinfo{journal}{J.
  Phys. Soc. Jpn.} \textbf{\bibinfo{volume}{67}}, \bibinfo{pages}{2421}
  (\bibinfo{year}{1998}).

\bibitem[{\citenamefont{Ando et~al.}(2002)\citenamefont{Ando, Zheng, and
  Suzuura}}]{Ando:jpsj02}
\bibinfo{author}{\bibfnamefont{T.}~\bibnamefont{Ando}},
  \bibinfo{author}{\bibfnamefont{Y.}~\bibnamefont{Zheng}}, \bibnamefont{and}
  \bibinfo{author}{\bibfnamefont{H.}~\bibnamefont{Suzuura}},
  \bibinfo{journal}{J. Phys. Soc. Jpn.} \textbf{\bibinfo{volume}{71}},
  \bibinfo{pages}{1318} (\bibinfo{year}{2002}).

\bibitem[{\citenamefont{Gusynin and Sharapov}(2005)}]{Gusynin:prl05}
\bibinfo{author}{\bibfnamefont{V.~P.} \bibnamefont{Gusynin}} \bibnamefont{and}
  \bibinfo{author}{\bibfnamefont{S.~G.} \bibnamefont{Sharapov}},
  \bibinfo{journal}{Phys. Rev. Lett.} \textbf{\bibinfo{volume}{95}},
  \bibinfo{pages}{146801} (\bibinfo{year}{2005}).

\bibitem[{\citenamefont{Katsnelson et~al.}(2006)\citenamefont{Katsnelson,
  Novoselov, and Geim}}]{Katsnelson:natp06}
\bibinfo{author}{\bibfnamefont{M.~I.} \bibnamefont{Katsnelson}},
  \bibinfo{author}{\bibfnamefont{K.~S.} \bibnamefont{Novoselov}},
  \bibnamefont{and} \bibinfo{author}{\bibfnamefont{A.~K.} \bibnamefont{Geim}},
  \bibinfo{journal}{Nature Physics} \textbf{\bibinfo{volume}{2}},
  \bibinfo{pages}{620} (\bibinfo{year}{2006}).

\bibitem[{\citenamefont{van~den Brink}(2007)}]{vandenBrink:natn07}
\bibinfo{author}{\bibfnamefont{J.}~\bibnamefont{van~den Brink}},
  \bibinfo{journal}{Nature Nanotechnology} \textbf{\bibinfo{volume}{2}},
  \bibinfo{pages}{199} (\bibinfo{year}{2007}).

\bibitem[{\citenamefont{Sabio et~al.}(2008)\citenamefont{Sabio, Seoanez,
  Fratini, Guinea, Neto, and Sols}}]{Sabio:prb08}
\bibinfo{author}{\bibfnamefont{J.}~\bibnamefont{Sabio}},
  \bibinfo{author}{\bibfnamefont{C.}~\bibnamefont{Seoanez}},
  \bibinfo{author}{\bibfnamefont{S.}~\bibnamefont{Fratini}},
  \bibinfo{author}{\bibfnamefont{F.}~\bibnamefont{Guinea}},
  \bibinfo{author}{\bibfnamefont{A.~H.~C.} \bibnamefont{Neto}},
  \bibnamefont{and} \bibinfo{author}{\bibfnamefont{F.}~\bibnamefont{Sols}},
  \bibinfo{journal}{Phys. Rev. B} \textbf{\bibinfo{volume}{77}},
  \bibinfo{pages}{195409} (\bibinfo{year}{2008}).

\bibitem[{\citenamefont{Boukhvalov et~al.}(2008)\citenamefont{Boukhvalov,
  Katsnelson, and Lichtenstein}}]{Boukhvalov:prb08}
\bibinfo{author}{\bibfnamefont{D.~W.} \bibnamefont{Boukhvalov}},
  \bibinfo{author}{\bibfnamefont{M.~I.} \bibnamefont{Katsnelson}},
  \bibnamefont{and} \bibinfo{author}{\bibfnamefont{A.~I.}
  \bibnamefont{Lichtenstein}}, \bibinfo{journal}{Phys. Rev. B}
  \textbf{\bibinfo{volume}{77}}, \bibinfo{pages}{035427}
  (\bibinfo{year}{2008}).

\bibitem[{\citenamefont{Lee et~al.}(2008)\citenamefont{Lee, Balasubramanian,
  Weitz, Burghard, and Kern}}]{Lee:natn08}
\bibinfo{author}{\bibfnamefont{E.~J.~H.} \bibnamefont{Lee}},
  \bibinfo{author}{\bibfnamefont{K.}~\bibnamefont{Balasubramanian}},
  \bibinfo{author}{\bibfnamefont{R.~T.} \bibnamefont{Weitz}},
  \bibinfo{author}{\bibfnamefont{M.}~\bibnamefont{Burghard}}, \bibnamefont{and}
  \bibinfo{author}{\bibfnamefont{K.}~\bibnamefont{Kern}},
  \bibinfo{journal}{Nature Nanotechnology} \textbf{\bibinfo{volume}{3}},
  \bibinfo{pages}{486} (\bibinfo{year}{2008}).

\bibitem[{\citenamefont{Oshima and Nagashima}(1997)}]{Oshima:jpcm97}
\bibinfo{author}{\bibfnamefont{C.}~\bibnamefont{Oshima}} \bibnamefont{and}
  \bibinfo{author}{\bibfnamefont{A.}~\bibnamefont{Nagashima}},
  \bibinfo{journal}{J. Phys.: Condens. Matter} \textbf{\bibinfo{volume}{9}},
  \bibinfo{pages}{1} (\bibinfo{year}{1997}).

\bibitem[{\citenamefont{Dedkov et~al.}(2001)\citenamefont{Dedkov, Shikin,
  Adamchuk, Molodtsov, Laubschat, Bauer, and Kaindl}}]{Dedkov:prb01}
\bibinfo{author}{\bibfnamefont{Y.~S.} \bibnamefont{Dedkov}},
  \bibinfo{author}{\bibfnamefont{A.~M.} \bibnamefont{Shikin}},
  \bibinfo{author}{\bibfnamefont{V.~K.} \bibnamefont{Adamchuk}},
  \bibinfo{author}{\bibfnamefont{S.~L.} \bibnamefont{Molodtsov}},
  \bibinfo{author}{\bibfnamefont{C.}~\bibnamefont{Laubschat}},
  \bibinfo{author}{\bibfnamefont{A.}~\bibnamefont{Bauer}}, \bibnamefont{and}
  \bibinfo{author}{\bibfnamefont{G.}~\bibnamefont{Kaindl}},
  \bibinfo{journal}{Phys. Rev. B} \textbf{\bibinfo{volume}{64}},
  \bibinfo{pages}{035405} (\bibinfo{year}{2001}).

\bibitem[{\citenamefont{Bertoni et~al.}(2005)\citenamefont{Bertoni, Calmels,
  Altibelli, and Serin}}]{Bertoni:prb05}
\bibinfo{author}{\bibfnamefont{G.}~\bibnamefont{Bertoni}},
  \bibinfo{author}{\bibfnamefont{L.}~\bibnamefont{Calmels}},
  \bibinfo{author}{\bibfnamefont{A.}~\bibnamefont{Altibelli}},
  \bibnamefont{and} \bibinfo{author}{\bibfnamefont{V.}~\bibnamefont{Serin}},
  \bibinfo{journal}{Phys. Rev. B} \textbf{\bibinfo{volume}{71}},
  \bibinfo{pages}{075402} (\bibinfo{year}{2005}).

\bibitem[{\citenamefont{{N'Diaye} et~al.}(2006)\citenamefont{{N'Diaye},
  Bleikamp, Feibelman, and Michely}}]{NDiaye:prl06}
\bibinfo{author}{\bibfnamefont{A.~T.} \bibnamefont{{N'Diaye}}},
  \bibinfo{author}{\bibfnamefont{S.}~\bibnamefont{Bleikamp}},
  \bibinfo{author}{\bibfnamefont{P.~J.} \bibnamefont{Feibelman}},
  \bibnamefont{and} \bibinfo{author}{\bibfnamefont{T.}~\bibnamefont{Michely}},
  \bibinfo{journal}{Phys. Rev. Lett.} \textbf{\bibinfo{volume}{97}},
  \bibinfo{pages}{215501} (\bibinfo{year}{2006}).

\bibitem[{\citenamefont{Karpan et~al.}(2007)\citenamefont{Karpan, Giovannetti,
  Khomyakov, Talanana, Starikov, Zwierzycki, van~den Brink, Brocks, and
  Kelly}}]{Karpan:prl07}
\bibinfo{author}{\bibfnamefont{V.~M.} \bibnamefont{Karpan}},
  \bibinfo{author}{\bibfnamefont{G.}~\bibnamefont{Giovannetti}},
  \bibinfo{author}{\bibfnamefont{P.~A.} \bibnamefont{Khomyakov}},
  \bibinfo{author}{\bibfnamefont{M.}~\bibnamefont{Talanana}},
  \bibinfo{author}{\bibfnamefont{A.~A.} \bibnamefont{Starikov}},
  \bibinfo{author}{\bibfnamefont{M.}~\bibnamefont{Zwierzycki}},
  \bibinfo{author}{\bibfnamefont{J.}~\bibnamefont{van~den Brink}},
  \bibinfo{author}{\bibfnamefont{G.}~\bibnamefont{Brocks}}, \bibnamefont{and}
  \bibinfo{author}{\bibfnamefont{P.~J.} \bibnamefont{Kelly}},
  \bibinfo{journal}{Phys. Rev. Lett.} \textbf{\bibinfo{volume}{99}},
  \bibinfo{pages}{176602} (\bibinfo{year}{2007}).

\bibitem[{\citenamefont{Giovannetti et~al.}(2007)\citenamefont{Giovannetti,
  Khomyakov, Brocks, Kelly, and van~den Brink}}]{Giovannetti:prb07}
\bibinfo{author}{\bibfnamefont{G.}~\bibnamefont{Giovannetti}},
  \bibinfo{author}{\bibfnamefont{P.~A.} \bibnamefont{Khomyakov}},
  \bibinfo{author}{\bibfnamefont{G.}~\bibnamefont{Brocks}},
  \bibinfo{author}{\bibfnamefont{P.~J.} \bibnamefont{Kelly}}, \bibnamefont{and}
  \bibinfo{author}{\bibfnamefont{J.}~\bibnamefont{van~den Brink}},
  \bibinfo{journal}{Phys. Rev. B} \textbf{\bibinfo{volume}{76}},
  \bibinfo{pages}{073103} (\bibinfo{year}{2007}).

\bibitem[{\citenamefont{Marchini et~al.}(2007)\citenamefont{Marchini,
  G{\"{u}}nther, and Wintterlin}}]{Marchini:prb07}
\bibinfo{author}{\bibfnamefont{S.}~\bibnamefont{Marchini}},
  \bibinfo{author}{\bibfnamefont{S.}~\bibnamefont{G{\"{u}}nther}},
  \bibnamefont{and}
  \bibinfo{author}{\bibfnamefont{J.}~\bibnamefont{Wintterlin}},
  \bibinfo{journal}{Phys. Rev. B} \textbf{\bibinfo{volume}{76}},
  \bibinfo{pages}{075429} (\bibinfo{year}{2007}).

\bibitem[{\citenamefont{Uchoa et~al.}(2008)\citenamefont{Uchoa, Lin, and
  Neto}}]{Uchoa:prb08}
\bibinfo{author}{\bibfnamefont{B.}~\bibnamefont{Uchoa}},
  \bibinfo{author}{\bibfnamefont{C.-Y.} \bibnamefont{Lin}}, \bibnamefont{and}
  \bibinfo{author}{\bibfnamefont{A.~H.~C.} \bibnamefont{Neto}},
  \bibinfo{journal}{Phys. Rev. B} \textbf{\bibinfo{volume}{77}},
  \bibinfo{pages}{035420} (\bibinfo{year}{2008}).

\bibitem[{\citenamefont{Rotenberg et~al.}(2008)\citenamefont{Rotenberg,
  Bostwick, Ohta, McChesney, Seyller, and Horn}}]{Rotenberg:natm08}
\bibinfo{author}{\bibfnamefont{E.}~\bibnamefont{Rotenberg}},
  \bibinfo{author}{\bibfnamefont{A.}~\bibnamefont{Bostwick}},
  \bibinfo{author}{\bibfnamefont{T.}~\bibnamefont{Ohta}},
  \bibinfo{author}{\bibfnamefont{J.~L.} \bibnamefont{McChesney}},
  \bibinfo{author}{\bibfnamefont{T.}~\bibnamefont{Seyller}}, \bibnamefont{and}
  \bibinfo{author}{\bibfnamefont{K.}~\bibnamefont{Horn}},
  \bibinfo{journal}{Nature Materials} \textbf{\bibinfo{volume}{7}},
  \bibinfo{pages}{258} (\bibinfo{year}{2008}).

\bibitem[{\citenamefont{Wu et~al.}(2008)\citenamefont{Wu, Sprinkle, Li, Ming,
  Berger, and de~Heer}}]{Wu:prl08}
\bibinfo{author}{\bibfnamefont{X.~S.} \bibnamefont{Wu}},
  \bibinfo{author}{\bibfnamefont{M.}~\bibnamefont{Sprinkle}},
  \bibinfo{author}{\bibfnamefont{X.~B.} \bibnamefont{Li}},
  \bibinfo{author}{\bibfnamefont{F.}~\bibnamefont{Ming}},
  \bibinfo{author}{\bibfnamefont{C.}~\bibnamefont{Berger}}, \bibnamefont{and}
  \bibinfo{author}{\bibfnamefont{W.~A.} \bibnamefont{de~Heer}},
  \bibinfo{journal}{Phys. Rev. Lett.} \textbf{\bibinfo{volume}{101}},
  \bibinfo{pages}{026801} (\bibinfo{year}{2008}).

\bibitem[{\citenamefont{Giovannetti et~al.}(2008)\citenamefont{Giovannetti,
  Khomyakov, Brocks, Karpan, van~den Brink, and Kelly}}]{Giovannetti:prl08}
\bibinfo{author}{\bibfnamefont{G.}~\bibnamefont{Giovannetti}},
  \bibinfo{author}{\bibfnamefont{P.~A.} \bibnamefont{Khomyakov}},
  \bibinfo{author}{\bibfnamefont{G.}~\bibnamefont{Brocks}},
  \bibinfo{author}{\bibfnamefont{V.~M.} \bibnamefont{Karpan}},
  \bibinfo{author}{\bibfnamefont{J.}~\bibnamefont{van~den Brink}},
  \bibnamefont{and} \bibinfo{author}{\bibfnamefont{P.~J.} \bibnamefont{Kelly}},
  \bibinfo{journal}{Phys. Rev. Lett.} \textbf{\bibinfo{volume}{101}},
  \bibinfo{pages}{026803} (\bibinfo{year}{2008}).

\bibitem[{\citenamefont{Schomerus}(2007)}]{Schomerus:prb07}
\bibinfo{author}{\bibfnamefont{H.}~\bibnamefont{Schomerus}},
  \bibinfo{journal}{Phys. Rev. B} \textbf{\bibinfo{volume}{76}},
  \bibinfo{pages}{045433} (\bibinfo{year}{2007}).

\bibitem[{\citenamefont{Blanter and Martin}(2007)}]{Blanter:prb07}
\bibinfo{author}{\bibfnamefont{Y.~M.} \bibnamefont{Blanter}} \bibnamefont{and}
  \bibinfo{author}{\bibfnamefont{I.}~\bibnamefont{Martin}},
  \bibinfo{journal}{Phys. Rev. B} \textbf{\bibinfo{volume}{76}},
  \bibinfo{pages}{155433} (\bibinfo{year}{2007}).
  
\bibitem{Huard:prb08}  B. Huard, N. Stander, J. A. Sulpizio, and D. Goldhaber-Gordon, Phys. Rev. B {\bf 78}, 121402(R) (2008).

\bibitem{Nouchi:apl08}  R. Nouchi, M. Shiraishi, and Y. Suzuki, Appl. Phys. Lett. {\bf 93}, 152104 (2008).

\bibitem{Russo:condmat09}  S. Russo, M. F. Craciun, M. Yamamoto, A. F. Morpurgo, and S. Tarucha, arXiv:0901.0485.

\bibitem[{\citenamefont{Cheianov and Fal'ko}(2006)}]{Cheianov:prb06}
\bibinfo{author}{\bibfnamefont{V.~V.} \bibnamefont{Cheianov}} \bibnamefont{and}
  \bibinfo{author}{\bibfnamefont{V.~I.} \bibnamefont{Fal'ko}},
  \bibinfo{journal}{Phys. Rev. B} \textbf{\bibinfo{volume}{74}},
  \bibinfo{pages}{041403} (\bibinfo{year}{2006}).

\bibitem[{\citenamefont{Huard et~al.}(2007)\citenamefont{Huard, Sulpizio,
  Stander, Todd, Yang, and Goldhaber-Gordon}}]{Huard:prl07}
\bibinfo{author}{\bibfnamefont{B.}~\bibnamefont{Huard}},
  \bibinfo{author}{\bibfnamefont{J.~A.} \bibnamefont{Sulpizio}},
  \bibinfo{author}{\bibfnamefont{N.}~\bibnamefont{Stander}},
  \bibinfo{author}{\bibfnamefont{K.}~\bibnamefont{Todd}},
  \bibinfo{author}{\bibfnamefont{B.}~\bibnamefont{Yang}}, \bibnamefont{and}
  \bibinfo{author}{\bibfnamefont{D.}~\bibnamefont{Goldhaber-Gordon}},
  \bibinfo{journal}{Phys. Rev. Lett.} \textbf{\bibinfo{volume}{98}},
  \bibinfo{pages}{236803} (\bibinfo{year}{2007}).

\bibitem[{\citenamefont{{\"{O}}zyilmaz
  et~al.}(2007)\citenamefont{{\"{O}}zyilmaz, Jarillo-Herrero, Efetov, Abanin,
  Levitov, and Kim}}]{Ozyilmaz:prl07}
\bibinfo{author}{\bibfnamefont{B.}~\bibnamefont{{\"{O}}zyilmaz}},
  \bibinfo{author}{\bibfnamefont{P.}~\bibnamefont{Jarillo-Herrero}},
  \bibinfo{author}{\bibfnamefont{D.}~\bibnamefont{Efetov}},
  \bibinfo{author}{\bibfnamefont{D.~A.} \bibnamefont{Abanin}},
  \bibinfo{author}{\bibfnamefont{L.~S.} \bibnamefont{Levitov}},
  \bibnamefont{and} \bibinfo{author}{\bibfnamefont{P.}~\bibnamefont{Kim}},
  \bibinfo{journal}{Phys. Rev. Lett.} \textbf{\bibinfo{volume}{99}},
  \bibinfo{pages}{166804} (\bibinfo{year}{2007}).

\bibitem[{\citenamefont{Williams et~al.}(2007)\citenamefont{Williams, DiCarlo,
  and Marcus}}]{Williams:sc07}
\bibinfo{author}{\bibfnamefont{J.~R.} \bibnamefont{Williams}},
  \bibinfo{author}{\bibfnamefont{L.}~\bibnamefont{DiCarlo}}, \bibnamefont{and}
  \bibinfo{author}{\bibfnamefont{C.~M.} \bibnamefont{Marcus}},
  \bibinfo{journal}{Science} \textbf{\bibinfo{volume}{317}},
  \bibinfo{pages}{638} (\bibinfo{year}{2007}).

\bibitem[{\citenamefont{Tworzyd{\l}o et~al.}(2007)\citenamefont{Tworzyd{\l}o,
  Snyman, Akhmerov, and Beenakker}}]{Tworzydlo:prb07}
\bibinfo{author}{\bibfnamefont{J.}~\bibnamefont{Tworzyd{\l}o}},
  \bibinfo{author}{\bibfnamefont{I.}~\bibnamefont{Snyman}},
  \bibinfo{author}{\bibfnamefont{A.~R.} \bibnamefont{Akhmerov}},
  \bibnamefont{and} \bibinfo{author}{\bibfnamefont{C.~W.~J.}
  \bibnamefont{Beenakker}}, \bibinfo{journal}{Phys. Rev. B}
  \textbf{\bibinfo{volume}{76}}, \bibinfo{pages}{035411}
  (\bibinfo{year}{2007}).

\bibitem[{\citenamefont{Fogler et~al.}(2008)\citenamefont{Fogler, Novikov,
  Glazman, and Shklovskii}}]{Fogler:prb08}
\bibinfo{author}{\bibfnamefont{M.~M.} \bibnamefont{Fogler}},
  \bibinfo{author}{\bibfnamefont{D.~S.} \bibnamefont{Novikov}},
  \bibinfo{author}{\bibfnamefont{L.~I.} \bibnamefont{Glazman}},
  \bibnamefont{and} \bibinfo{author}{\bibfnamefont{B.~I.}
  \bibnamefont{Shklovskii}}, \bibinfo{journal}{Phys. Rev. B}
  \textbf{\bibinfo{volume}{77}}, \bibinfo{pages}{075420}
  (\bibinfo{year}{2008}).

\bibitem[{\citenamefont{Park et~al.}(2008)\citenamefont{Park, Yang, So, Cohen,
  and Louie}}]{Park:natp08}
\bibinfo{author}{\bibfnamefont{C.-H.} \bibnamefont{Park}},
  \bibinfo{author}{\bibfnamefont{L.}~\bibnamefont{Yang}},
  \bibinfo{author}{\bibfnamefont{Y.-W.} \bibnamefont{So}},
  \bibinfo{author}{\bibfnamefont{M.~L.} \bibnamefont{Cohen}}, \bibnamefont{and}
  \bibinfo{author}{\bibfnamefont{S.~G.} \bibnamefont{Louie}},
  \bibinfo{journal}{Nature Physics} \textbf{\bibinfo{volume}{4}},
  \bibinfo{pages}{213} (\bibinfo{year}{2008}).

\bibitem[{\citenamefont{Gorbachev et~al.}(2008)\citenamefont{Gorbachev,
  Mayorov, Savchenko, Horsell, and Guinea}}]{Gorbachev:nanol08}
\bibinfo{author}{\bibfnamefont{R.~V.} \bibnamefont{Gorbachev}},
  \bibinfo{author}{\bibfnamefont{A.~S.} \bibnamefont{Mayorov}},
  \bibinfo{author}{\bibfnamefont{A.~K.} \bibnamefont{Savchenko}},
  \bibinfo{author}{\bibfnamefont{D.~W.} \bibnamefont{Horsell}},
  \bibnamefont{and} \bibinfo{author}{\bibfnamefont{F.}~\bibnamefont{Guinea}},
  \bibinfo{journal}{Nano Letters} \textbf{\bibinfo{volume}{8}},
  \bibinfo{pages}{1995} (\bibinfo{year}{2008}).

\bibitem[{\citenamefont{Okada and Oshiyama}(2005)}]{Okada:prl05}
\bibinfo{author}{\bibfnamefont{S.}~\bibnamefont{Okada}} \bibnamefont{and}
  \bibinfo{author}{\bibfnamefont{A.}~\bibnamefont{Oshiyama}},
  \bibinfo{journal}{Phys. Rev. Lett.} \textbf{\bibinfo{volume}{95}},
  \bibinfo{pages}{206804} (\bibinfo{year}{2005}).

\bibitem[{\citenamefont{Nosho et~al.}(2005)\citenamefont{Nosho, Ohno,
  Kishimoto, and Mizutani}}]{Nosho:apl05}
\bibinfo{author}{\bibfnamefont{Y.}~\bibnamefont{Nosho}},
  \bibinfo{author}{\bibfnamefont{Y.}~\bibnamefont{Ohno}},
  \bibinfo{author}{\bibfnamefont{S.}~\bibnamefont{Kishimoto}},
  \bibnamefont{and} \bibinfo{author}{\bibfnamefont{T.}~\bibnamefont{Mizutani}},
  \bibinfo{journal}{Appl. Phys. Lett.} \textbf{\bibinfo{volume}{86}},
  \bibinfo{pages}{073105} (\bibinfo{year}{2005}).

\bibitem[{\citenamefont{Zhu and Kaxiras}(2006)}]{Zhu:apl06}
\bibinfo{author}{\bibfnamefont{W.}~\bibnamefont{Zhu}} \bibnamefont{and}
  \bibinfo{author}{\bibfnamefont{E.}~\bibnamefont{Kaxiras}},
  \bibinfo{journal}{Appl. Phys. Lett.} \textbf{\bibinfo{volume}{89}},
  \bibinfo{pages}{243107} (\bibinfo{year}{2006}).

\bibitem[{\citenamefont{Meng et~al.}(2007)\citenamefont{Meng, Wang, and
  Wang}}]{Meng:jap07}
\bibinfo{author}{\bibfnamefont{T.~Z.} \bibnamefont{Meng}},
  \bibinfo{author}{\bibfnamefont{C.-Y.} \bibnamefont{Wang}}, \bibnamefont{and}
  \bibinfo{author}{\bibfnamefont{S.-Y.} \bibnamefont{Wang}},
  \bibinfo{journal}{J. Appl. Phys.} \textbf{\bibinfo{volume}{102}},
  \bibinfo{pages}{013709} (\bibinfo{year}{2007}).
  
\bibitem{Vitale:jacs08} V. Vitale, A. Curioni, and W. Andreoni, JACS {\bf 130}, 5848 (2008).

\bibitem[{\citenamefont{Khomyakov
  et~al.}(2009{\natexlab{a}})\citenamefont{Khomyakov, Giovannetti, Rusu,
  Brocks, van~den Brink, and Kelly}}]{Khomyakov:unpub1}
\bibinfo{author}{\bibfnamefont{P.~A.} \bibnamefont{Khomyakov}},
  \bibinfo{author}{\bibfnamefont{G.}~\bibnamefont{Giovannetti}},
  \bibinfo{author}{\bibfnamefont{P.~C.} \bibnamefont{Rusu}},
  \bibinfo{author}{\bibfnamefont{G.}~\bibnamefont{Brocks}},
  \bibinfo{author}{\bibfnamefont{J.}~\bibnamefont{van~den Brink}},
  \bibnamefont{and} \bibinfo{author}{\bibfnamefont{P.~J.} \bibnamefont{Kelly}}
  (\bibinfo{year}{2009}{\natexlab{a}}), \bibinfo{note}{to be published}.

\bibitem[{\citenamefont{Jonker et~al.}(1981)\citenamefont{Jonker, Morar, and
  Park}}]{Jonker:prb81}
\bibinfo{author}{\bibfnamefont{B.~T.} \bibnamefont{Jonker}},
  \bibinfo{author}{\bibfnamefont{J.~F.} \bibnamefont{Morar}}, \bibnamefont{and}
  \bibinfo{author}{\bibfnamefont{R.~L.} \bibnamefont{Park}},
  \bibinfo{journal}{Phys. Rev. B} \textbf{\bibinfo{volume}{24}},
  \bibinfo{pages}{2951} (\bibinfo{year}{1981}).

\bibitem[{\citenamefont{Michaelson}(1977)}]{Michaelson:jap77}
\bibinfo{author}{\bibfnamefont{H.~B.} \bibnamefont{Michaelson}},
  \bibinfo{journal}{J. Appl. Phys.} \textbf{\bibinfo{volume}{48}},
  \bibinfo{pages}{4729} (\bibinfo{year}{1977}).

\bibitem[{\citenamefont{Vaara et~al.}(1993)\citenamefont{Vaara, Vaari, and
  Lahtinen}}]{Vaara:ss98}
\bibinfo{author}{\bibfnamefont{T.}~\bibnamefont{Vaara}},
  \bibinfo{author}{\bibfnamefont{J.}~\bibnamefont{Vaari}}, \bibnamefont{and}
  \bibinfo{author}{\bibfnamefont{J.}~\bibnamefont{Lahtinen}},
  \bibinfo{journal}{Surface Science} \textbf{\bibinfo{volume}{395}},
  \bibinfo{pages}{88} (\bibinfo{year}{1993}).

\bibitem[{\citenamefont{Derry and Ji-Zhong}(1989)}]{Derry:prb89}
\bibinfo{author}{\bibfnamefont{G.~N.} \bibnamefont{Derry}} \bibnamefont{and}
  \bibinfo{author}{\bibfnamefont{Z.}~\bibnamefont{Ji-Zhong}},
  \bibinfo{journal}{Phys. Rev. B} \textbf{\bibinfo{volume}{39}},
  \bibinfo{pages}{1940} (\bibinfo{year}{1989}).

\bibitem[{\citenamefont{Perdew and Zunger}(1981)}]{Perdew:prb81}
\bibinfo{author}{\bibfnamefont{J.~P.} \bibnamefont{Perdew}} \bibnamefont{and}
  \bibinfo{author}{\bibfnamefont{A.}~\bibnamefont{Zunger}},
  \bibinfo{journal}{Phys. Rev. B} \textbf{\bibinfo{volume}{23}},
  \bibinfo{pages}{5048} (\bibinfo{year}{1981}).

\bibitem[{\citenamefont{Bl{\"{o}}chl}(1994)}]{Blochl:prb94b}
\bibinfo{author}{\bibfnamefont{P.~E.} \bibnamefont{Bl{\"{o}}chl}},
  \bibinfo{journal}{Phys. Rev. B} \textbf{\bibinfo{volume}{50}},
  \bibinfo{pages}{17953} (\bibinfo{year}{1994}).

\bibitem[{\citenamefont{Kresse and Joubert}(1999)}]{Kresse:prb99}
\bibinfo{author}{\bibfnamefont{G.}~\bibnamefont{Kresse}} \bibnamefont{and}
  \bibinfo{author}{\bibfnamefont{D.}~\bibnamefont{Joubert}},
  \bibinfo{journal}{Phys. Rev. B} \textbf{\bibinfo{volume}{59}},
  \bibinfo{pages}{1758} (\bibinfo{year}{1999}).

\bibitem[{\citenamefont{Kresse and Hafner}(1993)}]{Kresse:prb93}
\bibinfo{author}{\bibfnamefont{G.}~\bibnamefont{Kresse}} \bibnamefont{and}
  \bibinfo{author}{\bibfnamefont{J.}~\bibnamefont{Hafner}},
  \bibinfo{journal}{Phys. Rev. B} \textbf{\bibinfo{volume}{47}},
  \bibinfo{pages}{558} (\bibinfo{year}{1993}).

\bibitem[{\citenamefont{Kresse and Furthmuller}(1996)}]{Kresse:prb96}
\bibinfo{author}{\bibfnamefont{G.}~\bibnamefont{Kresse}} \bibnamefont{and}
  \bibinfo{author}{\bibfnamefont{J.}~\bibnamefont{Furthmuller}},
  \bibinfo{journal}{Phys. Rev. B} \textbf{\bibinfo{volume}{54}},
  \bibinfo{pages}{11169} (\bibinfo{year}{1996}).

\bibitem[{\citenamefont{Neugebauer and Scheffler}(1992)}]{Neugebauer:prb92}
\bibinfo{author}{\bibfnamefont{J.}~\bibnamefont{Neugebauer}} \bibnamefont{and}
  \bibinfo{author}{\bibfnamefont{M.}~\bibnamefont{Scheffler}},
  \bibinfo{journal}{Phys. Rev. B} \textbf{\bibinfo{volume}{46}},
  \bibinfo{pages}{16067} (\bibinfo{year}{1992}).

\bibitem[{\citenamefont{Bl{\"{o}}chl et~al.}(1994)\citenamefont{Bl{\"{o}}chl,
  Jepsen, and Andersen}}]{Blochl:prb94a}
\bibinfo{author}{\bibfnamefont{P.~E.} \bibnamefont{Bl{\"{o}}chl}},
  \bibinfo{author}{\bibfnamefont{O.}~\bibnamefont{Jepsen}}, \bibnamefont{and}
  \bibinfo{author}{\bibfnamefont{O.~K.} \bibnamefont{Andersen}},
  \bibinfo{journal}{Phys. Rev. B} \textbf{\bibinfo{volume}{49}},
  \bibinfo{pages}{16223} (\bibinfo{year}{1994}).

\bibitem[{\citenamefont{Gamo et~al.}(1997)\citenamefont{Gamo, Nagashima,
  Wakabayashi, Terai, and Oshima}}]{Gamo:ss97}
\bibinfo{author}{\bibfnamefont{Y.}~\bibnamefont{Gamo}},
  \bibinfo{author}{\bibfnamefont{A.}~\bibnamefont{Nagashima}},
  \bibinfo{author}{\bibfnamefont{M.}~\bibnamefont{Wakabayashi}},
  \bibinfo{author}{\bibfnamefont{M.}~\bibnamefont{Terai}}, \bibnamefont{and}
  \bibinfo{author}{\bibfnamefont{C.}~\bibnamefont{Oshima}},
  \bibinfo{journal}{Surface Science} \textbf{\bibinfo{volume}{374}},
  \bibinfo{pages}{61} (\bibinfo{year}{1997}).

\bibitem[{\citenamefont{Qi et~al.}(2005)\citenamefont{Qi, {Hector Jr.}, Ooi,
  and B.Adams}}]{Qi:ss05}
\bibinfo{author}{\bibfnamefont{Y.}~\bibnamefont{Qi}},
  \bibinfo{author}{\bibfnamefont{L.~G.} \bibnamefont{{Hector Jr.}}},
  \bibinfo{author}{\bibfnamefont{N.}~\bibnamefont{Ooi}}, \bibnamefont{and}
  \bibinfo{author}{\bibfnamefont{J.}~\bibnamefont{B.Adams}},
  \bibinfo{journal}{Surface Science} \textbf{\bibinfo{volume}{581}},
  \bibinfo{pages}{155} (\bibinfo{year}{2005}).

\bibitem[{foo({\natexlab{a}})}]{footnote1}
\bibinfo{note}{The definition of $q$ in Eq.~\eqref{eq3} is somewhat arbitrary
  since $\Delta n(z)$ is a continuous function. It is used here only as an
  estimate for the interface dipole charge and $q$ should not be confused with
  the charge $eN$ transfered between the graphene sheet and the metal surface
  which is uniquely defined in terms of $\Delta E_{\rm F}$. For the metal
  surfaces studied in this paper, $q$ is always positive which leads to a
  decrease of the work function as compared to the clean metal surface. In
  contrast, $eN$, which determines the type of doping of graphene can be both
  positive and negative. The two charges, $q$ and $eN$, are related via
  Eq.~(\ref{eq:all}) since $eN\propto \Delta E_{\rm F}$ and $q\propto \Delta
  V$. One has $q=eN/N_{\rm C}$ only if $\Delta_{\rm c}(d)=0$, which holds if
  $d\gg d_{\rm eq}$, i.e., if the graphene sheet is far from the metal
  surface.}

\bibitem[{\citenamefont{Silva et~al.}(2003)\citenamefont{Silva, Stampf, and
  Scheffler}}]{Silva:prl03}
\bibinfo{author}{\bibfnamefont{J.~L.~F.} \bibnamefont{Silva}},
  \bibinfo{author}{\bibfnamefont{C.}~\bibnamefont{Stampf}}, \bibnamefont{and}
  \bibinfo{author}{\bibfnamefont{M.}~\bibnamefont{Scheffler}},
  \bibinfo{journal}{Phys. Rev. Lett.} \textbf{\bibinfo{volume}{90}},
  \bibinfo{pages}{066104} (\bibinfo{year}{2003}).

\bibitem[{\citenamefont{Rusu}(2007)}]{Rusu:thesis07}
\bibinfo{author}{\bibfnamefont{P.~C.} \bibnamefont{Rusu}}, Ph.D. thesis,
  \bibinfo{school}{University of Twente} (\bibinfo{year}{2007}),
  \bibinfo{note}{http://purl.org/utwente/58034}.

\bibitem[{foo({\natexlab{b}})}]{footnote2}
\bibinfo{note}{We obtain $\Delta_{\rm c}(d)$ by least-squares fitting
  Eq.~(\ref{eqn1}) to the DFT/LDA results for $\Delta E_{\rm F}(d)$ for Cu
  (111) with $d_0=2.4$ \AA. This value of $d_0$ provides the best fit of
  $\Delta E_{\rm F} (d)$. At large $d$ the chemical interaction term
  $\Delta_c(d)$ should vanish. Parametrizing $\Delta_{\rm c}(d)=e^{-\gamma d}\,
  (a_0 + a_1 d + a_2 d^2)$ yields $\gamma = 1.6443$ \AA$^{-1}$, $a_0 =
  -2048.56$ eV, $a_1 = 1363.87$ eV/\AA, and $a_2 = -205.737$ eV/\AA$^2$, where
  points with $d \gtrsim 3.0\,{\rm \AA}$ have been used for the fit. The
  general applicability of this equation to all metal substrates can be
  explained in terms of the weak metal-graphene interaction. The electron
  redistribution is then dominated by exchange repulsion and is almost
  independent of the metal species [\onlinecite{Rusu:thesis07}].}

\bibitem[{foo({\natexlab{c}})}]{footnote3}
\bibinfo{note}{The situation is analogous to what happens at the boundary
  between different facets of a metal crystal when the work function depends on
  the facet orientation. As a result of the difference in work functions, a
  transient electron current must flow from the lower workfunction to the
  higher workfunction surface. This charge rearrangement gives rise to a
  non-vanishing electrostatic potential field in the region of the boundary
  between the different surfaces so that there is no longer a well defined
  vacuum level.}

\bibitem[{\citenamefont{Perdew et~al.}(1992)\citenamefont{Perdew, Chevary,
  Vosko, Jackson, Pederson, Singh, and Fiolhais}}]{Perdew:prb92}
\bibinfo{author}{\bibfnamefont{J.~P.} \bibnamefont{Perdew}},
  \bibinfo{author}{\bibfnamefont{J.~A.} \bibnamefont{Chevary}},
  \bibinfo{author}{\bibfnamefont{S.~H.} \bibnamefont{Vosko}},
  \bibinfo{author}{\bibfnamefont{K.~A.} \bibnamefont{Jackson}},
  \bibinfo{author}{\bibfnamefont{M.~R.} \bibnamefont{Pederson}},
  \bibinfo{author}{\bibfnamefont{D.~J.} \bibnamefont{Singh}}, \bibnamefont{and}
  \bibinfo{author}{\bibfnamefont{C.}~\bibnamefont{Fiolhais}},
  \bibinfo{journal}{Phys. Rev. B} \textbf{\bibinfo{volume}{46}},
  \bibinfo{pages}{6671} (\bibinfo{year}{1992}).

\bibitem[{\citenamefont{Perdew et~al.}(1993)\citenamefont{Perdew, Chevary,
  Vosko, Jackson, Pederson, Singh, and Fiolhais}}]{Perdew:prb93}
\bibinfo{author}{\bibfnamefont{J.~P.} \bibnamefont{Perdew}},
  \bibinfo{author}{\bibfnamefont{J.~A.} \bibnamefont{Chevary}},
  \bibinfo{author}{\bibfnamefont{S.~H.} \bibnamefont{Vosko}},
  \bibinfo{author}{\bibfnamefont{K.~A.} \bibnamefont{Jackson}},
  \bibinfo{author}{\bibfnamefont{M.~R.} \bibnamefont{Pederson}},
  \bibinfo{author}{\bibfnamefont{D.~J.} \bibnamefont{Singh}}, \bibnamefont{and}
  \bibinfo{author}{\bibfnamefont{C.}~\bibnamefont{Fiolhais}},
  \bibinfo{journal}{Phys. Rev. B} \textbf{\bibinfo{volume}{48}},
  \bibinfo{pages}{4978} (\bibinfo{year}{1993}).

\bibitem[{\citenamefont{Rydberg et~al.}(2003)\citenamefont{Rydberg, Dion,
  Jacobson, Schroder, Hyldgaard, Simak, Langreth, and
  Lundqvist}}]{Rydberg:prl03}
\bibinfo{author}{\bibfnamefont{H.}~\bibnamefont{Rydberg}},
  \bibinfo{author}{\bibfnamefont{M.}~\bibnamefont{Dion}},
  \bibinfo{author}{\bibfnamefont{N.}~\bibnamefont{Jacobson}},
  \bibinfo{author}{\bibfnamefont{E.}~\bibnamefont{Schroder}},
  \bibinfo{author}{\bibfnamefont{P.}~\bibnamefont{Hyldgaard}},
  \bibinfo{author}{\bibfnamefont{S.~I.} \bibnamefont{Simak}},
  \bibinfo{author}{\bibfnamefont{D.~C.} \bibnamefont{Langreth}},
  \bibnamefont{and} \bibinfo{author}{\bibfnamefont{B.~I.}
  \bibnamefont{Lundqvist}}, \bibinfo{journal}{Phys. Rev. Lett.}
  \textbf{\bibinfo{volume}{91}}, \bibinfo{pages}{126402}
  (\bibinfo{year}{2003}).

\bibitem[{\citenamefont{Thonhauser et~al.}(2007)\citenamefont{Thonhauser,
  Cooper, Li, Puzder, Hyldgaard, and Langreth}}]{Thonhauser:prb07}
\bibinfo{author}{\bibfnamefont{T.}~\bibnamefont{Thonhauser}},
  \bibinfo{author}{\bibfnamefont{V.~R.} \bibnamefont{Cooper}},
  \bibinfo{author}{\bibfnamefont{S.}~\bibnamefont{Li}},
  \bibinfo{author}{\bibfnamefont{A.}~\bibnamefont{Puzder}},
  \bibinfo{author}{\bibfnamefont{P.}~\bibnamefont{Hyldgaard}},
  \bibnamefont{and} \bibinfo{author}{\bibfnamefont{D.~C.}
  \bibnamefont{Langreth}}, \bibinfo{journal}{Phys. Rev. B}
  \textbf{\bibinfo{volume}{76}}, \bibinfo{pages}{125112}
  (\bibinfo{year}{2007}).

\bibitem[{\citenamefont{Khomyakov
  et~al.}(2009{\natexlab{b}})\citenamefont{Khomyakov, Starikov, Brocks, and Kelly}}]{Khomyakov:unpub2}
\bibinfo{author}{\bibfnamefont{P.~A.} \bibnamefont{Khomyakov}},
  \bibinfo{author}{\bibfnamefont{A.~A.}~\bibnamefont{Starikov}},
  \bibinfo{author}{\bibfnamefont{G.}~\bibnamefont{Brocks}},
    \bibnamefont{and} \bibinfo{author}{\bibfnamefont{P.~J.} \bibnamefont{Kelly}}
  (\bibinfo{year}{2009}{\natexlab{b}}), \bibinfo{note}{to be published}.

\end{thebibliography}

\end{document}